\documentclass[twocolumn]{aastex701}

\usepackage{subcaption}
\usepackage{threeparttable}
\RequirePackage{amsmath}

\begin{document}

\title{The Disk Luminosity Deficit as a Tracer of Disk Recession during Soft-to-Hard Transitions in Black Hole X-ray Binaries}

\author[0009-0007-6828-3931]{Sai-En Xu}
\affiliation{Department of Astronomy, School of Physics, Wuhan University, Wuhan 430072, People's Republic of China}
\email{saienxu@whu.edu.cn}

\correspondingauthor{Bei You, Zhen Yan}
\author[0000-0002-8231-063X]{Bei You}
\affiliation{Department of Astronomy, School of Physics, Wuhan University, Wuhan 430072, People's Republic of China}
\email{youbei@whu.edu.cn}

\author[0009-0007-1966-181X]{Yi Long}
\affiliation{Department of Astronomy, Nanjing University, 163 Xianlin Avenue, Nanjing 210023, People's Republic of China}
\email{longyi@smail.nju.edu.cn}

\author[0000-0002-5385-9586]{Zhen Yan}
\affiliation{Shanghai Astronomical Observatory, Chinese Academy of Sciences (CAS), Shanghai 200030, China}
\email{zyan@shao.ac.cn}

\author[0000-0002-0333-2452]{Andrzej Zdziarski}
\affiliation{Nicolaus Copernicus Astronomical Center, Polish Academy of Sciences, Bartycka 18, Warszawa, PL-00-716, Poland}
\email{aaz@camk.edu.pl}

\author[0009-0002-6492-2920]{Rui-Xiang Hu}
\affiliation{Department of Astronomy, School of Physics, Wuhan University, Wuhan 430072, People's Republic of China}
\email{huruixiang@whu.edu.cn}

\author[0000-0002-2173-0673]{Alex Markowitz}
\affiliation{Nicolaus Copernicus Astronomical Center, Polish Academy of Sciences, Bartycka 18, Warszawa, PL-00-716, Poland}
\email{almarkowitz@camk.edu.pl}

\begin{abstract}

Tracing the evolution of the thin accretion disk during the soft-to-hard state transition in black hole X-ray binaries (BHXRBs) remains difficult because conventional spectral estimates of the disk inner radius become highly model-dependent once the thermal component weakens. As an alternative approach, we systematically analyze \textit{RXTE}/PCA observations of 26 BHXRBs and use the thermal disk luminosity deficit as an empirical tracer of disk recession. In 24 outbursts, the disk luminosity decays exponentially in the soft state, then drops significantly below the extrapolated baseline. This thermal luminosity deficit provides evidence for reduced disk radiative efficiency caused by the outward recession of the optically thick disk. Under this framework, we found that the estimated characteristic truncation radius increases rapidly as the systems evolve through the soft-to-hard transition. This interpretation is supported by timing analysis: in observations with well-constrained power density spectra, the characteristic frequencies of broadband noise and low-frequency QPOs generally decrease as the inferred truncation radius increases, consistent with the expansion of a hot inner flow.
The onset and rate of recession vary substantially across different sources and outbursts. Our results demonstrate that luminosity deficits provide a practical empirical tracer of thin-disk recession during soft-to-hard transitions, when direct spectral radius measurements become unreliable.

\end{abstract}


\section{Introduction} \label{sec:intro}

Low-mass black hole X-ray binaries (BHXRBs) spend most of their lifetimes in quiescence, but occasionally undergo bright outbursts during which the accretion rate increases by several orders of magnitude \citep{2006ARA&A..44...49R}. During an outburst, they typically evolve through a sequence of spectral states and trace a hysteretic track in the hardness--intensity diagram \citep{2005Ap&SS.300..107H, 2016ASSL..440...61B}. This spectral evolution is commonly interpreted as reflecting changes in the accretion geometry. In the truncated disk scenario, an outer geometrically thin, optically thick disk is replaced at inner radii by a hot inner flow; as the source decays back to the hard state, the inner edge of the disk is expected to recede to larger radii \citep{1997ApJ...489..865E, 2018A&A...614A..79P, 2022iSci...25j3544L}. Determining when and how this recession occurs remains an observational challenge.

Direct estimates of the truncation radius from X-ray spectra are difficult during the soft-to-hard transition. Although continuum fitting is widely used to infer the inner disk radius \citep[e.g.,][]{2008MNRAS.388..753G, 2014A&A...564A..37P, 2022ApJ...928...11Z}, the result becomes increasingly uncertain during the soft-to-hard transition, when the disk component weakens and becomes difficult to separate from the dominant Comptonized emission. In this regime, the disk temperature, normalization, absorption column, and coronal continuum are strongly coupled, while the conversion from the apparent radius to the physical radius depends further on the color correction factor $f_{\rm col}$ \citep{1995ApJ...445..780S, 2000MNRAS.313..193M}. Possible variations of $f_{\rm col}$ with temperature and accretion state add an additional source of uncertainty \citep{konig2026systematic}. These difficulties motivate methods that do not rely solely on direct spectral parameter estimates.

An alternative approach is to use the evolution of the thermal disk luminosity during the outburst decay. In irradiated disk models, X-ray irradiation from the central source modifies the thermal equilibrium of the accretion disk and can keep the outer disk in a hot, ionized state during the decay phase \citep{1990ApJ...359..164T}. The disk can then drain viscously, producing an approximately exponential decline in the light curve \citep{1998MNRAS.293L..42K}. The decay morphology is therefore connected to basic disk properties, such as the disk size, the mass inflow rate, and the viscosity parameter $\alpha$ \citep[e.g.,][]{1998MNRAS.301..382S, 2007MNRAS.374..466P, 2018Natur.554...69T}. In this context, deviations from the exponential trend established during soft state decay may carry information about changes in the accretion efficiency or geometry of the inner accretion flow.

\citet{2023Sci...381..961Y} reported a significant drop in the disk luminosity below the exponential trend during the soft-to-hard transition of MAXI J1820+070, accompanied by a Compton flare. They interpreted this behavior as evidence for disk recession and expansion of the hot inner flow, within a broader scenario that may lead to the formation of a magnetically arrested disk. Since the disk luminosity ($L_{\rm d}$) depends on the gravitational energy released near the inner boundary, with $L_{\rm d}\propto \dot{M}/R_{\rm tr}$ (where $\dot{M}$ is the mass accretion rate and $R_{\rm tr}$ is the truncation radius), an outward recession of the disk can reduce the disk accretion efficiency. The decay profile of the disk luminosity, therefore, provides a potential way to infer the evolution of the accretion geometry without relying directly on the apparent spectral radius.

Timing analysis can provide an independent consistency check on this interpretation. The power density spectra of BHXRBs commonly show broadband noise and low-frequency quasi-periodic oscillations (QPOs), whose characteristic frequencies are often linked to the size and timescales of the inner accretion flow \citep{1999ApJ...514..939W, 2002ApJ...572..392B}. In particular, the noise break frequency has been associated with the viscous timescale near the disk truncation radius \citep{1997MNRAS.292..679L, 2002ApJ...572..392B, 2011MNRAS.415.2323I}. More generally, recent reverberation studies of broadband noise further demonstrate that rapid variability can provide information on the disk--corona geometry \citep{2025ApJ...985..258Z, 2026NatCo..17.2860Y}. Low-frequency QPOs are frequently interpreted in terms of changes in the accretion flow geometry, such as Lense--Thirring precession \citep{2009MNRAS.397L.101I,2019NewAR..8501524I}. Thus, a decrease in these timing frequencies as the disk recedes would be qualitatively consistent with the expansion of the hot inner flow.

In this work, we systematically analyzed archival \textit{RXTE}/PCA observations of transient black-hole low-mass X-ray binaries (LMXBs). The parent sample was based on the black-hole LMXB catalog of \citet{2016A&A...587A..61C}, restricted to sources with well-sampled PCA coverage during one or more outburst decays. We excluded persistent or non-outbursting systems, neutron-star X-ray binaries, high-mass X-ray binaries, and sources for which the outburst was not adequately covered. This parent sample was used for spectral decomposition, from which we further selected episodes that exhibit a well-defined soft state disk decay and a subsequent thermal luminosity deficit. For these selected episodes, we isolated the disk component and used its deviation from an exponential baseline to construct a model-defined tracer of disk recession. We also extracted timing features to examine whether their evolution is consistent with the inferred geometric changes.

The structure of this paper is as follows. In Section~\ref{sec:data}, we describe the data reduction, spectral fitting, and timing analysis procedures. Section~\ref{sec:analytical_method} describes the analytical framework used to construct the disk recession tracer. In Section~\ref{sec:results}, we present the results of our spectral and timing analysis, focusing on the disk truncation radius inferred from the light-curve deviations. In Section~\ref{sec:discussion}, we discuss the diversity of disk recession across outbursts and compare our tracer with direct spectral inner-radius constraints. We also examine the roles of outer-disk irradiation, inner-disk evaporation, and disk radiative efficiency. Finally, a brief summary is given in Section~\ref{sec:conclusion}.

\section{Data Reduction}\label{sec:data}

\subsection{RXTE observations}

We performed a systematic analysis of archival data from the \textit{Rossi X-ray Timing Explorer (RXTE)}. We focused on the data obtained with the Proportional Counter Array (PCA). We employed the \texttt{maketime} task to generate Good Time Intervals (GTIs) and the \texttt{pcaextspec2} task to extract source and background spectra for each observation. Table~\ref{tab:sourcesinfo} summarizes the parent sample analyzed in this work, i.e., all BHXRB outbursts for which archival \textit{RXTE}/PCA spectra were extracted and fitted. The fundamental parameters of these sources were adopted from the catalogs of \citet{2016A&A...587A..61C} and \citet{2023A&A...675A.199A}. The final subset was selected after spectral fitting, based on the detectability of the disk component and the presence of a measurable deviation from the soft state exponential trend.

\def\tabOneCiteCorral{\citet{2016A&A...587A..61C}}
\def\tabOneCiteAvakyan{\citet{2023A&A...675A.199A}}
\def\tabOneCiteHone{\citep[e.g.,][]{2020A&A...637A..47A, 2023ApJ...943..165S}}
\def\tabOneCiteXone{\citep[e.g.,][]{2002A&A...390..199B, 2005ApJ...625..923S}}
\begin{deluxetable*}{lcccccc}
\tablewidth{0pt}
\tablecaption{Fundamental parameters of the BHXRB sample\label{tab:sourcesinfo}}
\tablehead{
    \colhead{Source} & \colhead{Outburst Year} & \colhead{$N_{\rm H}$} & \colhead{$M_{\rm BH}$} & \colhead{Distance} & \colhead{$P_{\rm orb}$} & \colhead{$i$} \\
    \colhead{} & \colhead{} & \colhead{($10^{22}$ cm$^{-2}$)} & \colhead{($M_{\odot}$)} & \colhead{(kpc)} & \colhead{(h)} & \colhead{($^\circ$)}
}
\startdata
4U 1543$-$47 & 2002 & 0.38 & $9.4 \pm 1.0$ & $7.95 \pm 1.45$ & $26.8$ & $20.7 \pm 1.5$ \\
GRO J1655$-$40 & 1996, 2005 & 0.74 & $6.0 \pm 0.4$ & $3.2 \pm 0.2$ & $62.9$ & $69 \pm 2$ \\
GS 1354$-$64 & 1997 & 0.8 & $7.6$ & $25$ & $61.1$ & $\le 79$ \\
GS 2023$+$338 & 2003 & 0.87 & $9.0^{+0.2}_{-0.6}$ & $2.39 \pm 0.14$ & $155.3$ & 52--80 \\
GX 339$-$4 & 2002, 2004, 2006, 2010 & 0.41 & $6.0 \pm 2.0$ & $7.0 \pm 2.0$ & $42.14 \pm 0.01$ & 37--78 \\
H1743$-$322 & 2002, other\tablenotemark{a} & 2.2 & $11.0 \pm 3.0$ & $10.0 \pm 3.0$ & -- & $75 \pm 3$ \\
IGR J17497$-$2821 & 2006 & 4.4 & -- & $8.0$ & -- & -- \\
MAXI J1543$-$564 & 2011 & 3.12 & -- & $10.75 \pm 2.25$ & -- & -- \\
MAXI J1659$-$152 & 2010 & 0.3 & $5.8 \pm 2.2$ & $6.95 \pm 1.65$ & $2.414 \pm 0.005$ & 65--80 \\
MAXI J1836$-$194 & 2011 & 0.19 & $7$ & $7 \pm 3$ & $< 4.9$ & 4--15 \\
SAX J1711.6$-$3808 & 2001 & 2.8 & -- & $8.5$ & -- & -- \\
SLX 1746$-$331 & 2002, 2003, 2007 & 1.77 & -- & $8.0$ & -- & -- \\
Swift J1357.2$-$0933 & 2011 & 0.01 & $12.4 \pm 3.6$ & $6.0$ & $2.8 \pm 0.3$ & $\gtrsim 80$ \\
Swift J1539.2$-$6227 & 2008 & 0.65 & -- & $8.0$ & -- & -- \\
Swift J1753.5$-$0127 & 2005--2011 & 0.23 & $8.8 \pm 1.3$ & $3.9 \pm 0.7$ & $3.26 \pm 0.02$ & -- \\
Swift J1842.5$-$1124 & 2008 & 0.36 & -- & $7 \pm 3$ & -- & -- \\
XTE J1118$+$480 & 1999, 2005 & 0.0074 & $7.2 \pm 0.72$ & $1.72 \pm 0.1$ & $4.08$ & 68--79 \\
XTE J1550$-$564 & 1998, 2000, other\tablenotemark{b} & 0.9 & $9.1 \pm 0.61$ & $4.38^{+0.58}_{-0.41}$ & $37.0$ & $75 \pm 4$ \\
XTE J1650$-$500 & 2001 & 0.53 & $7.3$ & $2.6 \pm 0.7$ & $7.69 \pm 0.02$ & $> 47$ \\
XTE J1652$-$453 & 2009 & 7.0 & $10.0$ & $8.0$ & -- & -- \\
XTE J1720$-$318 & 2003 & 1.3 & -- & $6.5 \pm 3.5$ & -- & -- \\
XTE J1748$-$288 & 1998 & 7.0 & -- & $8.0$ & -- & -- \\
XTE J1752$-$223 & 2009 & 0.86 & $9.5 \pm 1.1$ & $6.0 \pm 2.0$ & $< 7$ & $\le 49$ \\
XTE J1817$-$330 & 2006 & 0.12 & $6.0^{+4.0}_{-2.0}$ & $5.5 \pm 4.5$ & -- & -- \\
XTE J1818$-$245 & 2005 & 0.54 & -- & $3.55 \pm 0.75$ & -- & -- \\
XTE J1859$+$226 & 1999 & 0.3 & $8.0 \pm 2.0$ & $12.5 \pm 1.5$ & $6.58 \pm 0.05$ & $66.6 \pm 4.3$ \\
\enddata
\tablenotetext{a}{H1743$-$322 showed a complete outburst in 2002. Between 2003 and 2011, it underwent nine additional, relatively weak outbursts \tabOneCiteHone.}
\tablenotetext{b}{XTE~J1550$-$564 underwent three mini-outbursts between 2001 and 2004 \tabOneCiteXone.}
\tablecomments{The fundamental source parameters are primarily adopted from the compilations of \tabOneCiteCorral\ and \tabOneCiteAvakyan. The original references for individual measurements are given therein. For sources without direct distance measurements, we adopt a fiducial distance of 8~kpc, as indicated in the table. For sources without direct black hole mass measurements, we adopt a fiducial mass of $M_{\rm BH} = 10 \pm 2\,M_{\odot}$ when computing Eddington-scaled quantities. Orbital periods ($P_{\rm orb}$) and inclinations ($i$) are likewise compiled from \tabOneCiteCorral, \tabOneCiteAvakyan, and the references therein.}
\end{deluxetable*}

\subsection{Model setting and spectral fits}

Considering the limited energy coverage of \textit{RXTE}/PCA, we analyzed the spectra in the 3--25~keV band using \texttt{XSPEC} version 12.13.0 \citep{1996ASPC..101...17A}. The continuum was modeled with a multicolor disk blackbody component, \texttt{diskbb} \citep{1984PASJ...36..741M, 1986ApJ...308..635M}, convolved with the thermal Comptonization model \texttt{thcomp} \citep{2020MNRAS.492.5234Z}. In this configuration, the \texttt{diskbb} component provides the seed thermal disk emission, while \texttt{thcomp} describes the fraction of disk photons Comptonized by the corona. For observations showing structured residuals around the Fe~K$\alpha$ line, we included a phenomenological Gaussian component to account for the iron line emission commonly observed in BHXRBs \citep{2022hxga.book....9K}. A systematic error of 1\% was added to all spectra before fitting.

We therefore adopted the spectral model
\begin{equation*}
    \texttt{tbabs} * (\texttt{thcomp} \otimes \texttt{diskbb} + \texttt{gaussian}) ,
\end{equation*}
where the Gaussian component was omitted when no evident Fe~K$\alpha$ residuals were present.

The \texttt{tbabs} component accounts for interstellar absorption \citep{2000ApJ...542..914W}, with the hydrogen column density $N_{\rm H}$ fixed at the values listed in Table~\ref{tab:sourcesinfo} for each source. In \texttt{thcomp}, the electron temperature was fixed at $kT_{\rm e}=150$~keV, because the high-energy cutoff cannot be reliably constrained with the 3--25~keV PCA spectra. This value is broadly consistent with the high electron temperatures commonly inferred in hard-state black hole systems \citep[e.g.,][]{2017MNRAS.472.4220B, 2026ApJ..1005L..78Z}. To check the dependence on this assumption, we repeated the fits with $kT_{\rm e}=50$ and 300~keV. The fitted disk parameters, covering fraction, and photon-index evolution are not significantly changed for the purposes of this work. An example of this comparison for 4U~1543$-$47 is shown in Appendix Figure~\ref{fig:4u1543_paras}. The seed photon spectrum is provided self-consistently by the input \texttt{diskbb} component. The line energy of the Gaussian component was fixed at 6.4~keV, while its width and normalization were allowed to vary when the component was included. Because our goal is to obtain continuum and disk flux measurements across a large \textit{RXTE}/PCA sample with a 3--25\,keV bandpass, we use this phenomenological Gaussian rather than full relativistic reflection modeling to account for the Fe~K$\alpha$ feature.

In the hard state, the thermal disk component is often weak in the PCA band, and its parameters can become poorly constrained. We therefore flagged weakly constrained disk measurements. A disk measurement was considered well constrained only when the fractional uncertainty of the inner disk temperature $T_{\rm in}$ was smaller than 0.5, and the fractional uncertainty of the \texttt{diskbb} normalization was smaller than 1.0. Measurements that do not satisfy both criteria are flagged as weakly constrained and are shown with translucent markers in the relevant figures.

Absorption features have been reported in some sources, most notably GRO~J1655$-$40 \citep{1999ApJ...520..776S}. We did not include source-specific absorption components in the baseline model, in order to maintain a uniform fitting procedure for the full sample. This simplification is partly motivated by the limited spectral resolution of \textit{RXTE}/PCA, which cannot resolve narrow, highly ionized absorption lines in detail. However, lower-ionization absorption could in principle introduce broader continuum curvature and affect the inferred disk parameters.

As a robustness check, we refitted the GRO~J1655$-$40 spectra showing the strongest residual structures by adding a phenomenological absorption edge to the continuum model, i.e.,
\texttt{tbabs*(edge*thcomp$\otimes$diskbb + gaussian)}.
The resulting fits give broadly consistent \texttt{diskbb} and \texttt{thcomp} parameters. The corresponding component fluxes are also largely unchanged. This suggests that the unmodeled absorption features are unlikely to drive the disk luminosity decay trends discussed in this work.

We estimated the unabsorbed model fluxes using the \texttt{cflux} convolution model in \texttt{XSPEC}. For each observation, the input \texttt{diskbb} flux, the total \texttt{thcomp}$\otimes$\texttt{diskbb} flux, and the Gaussian line flux were evaluated separately over 0.01--1000~keV using the best-fit model constrained by the PCA spectra. This energy range is used to define a consistent model flux rather than a strictly bolometric luminosity, since the disk emission below the PCA band and the Comptonized emission above the PCA band are both model-dependent extrapolations. Following the definition in \citet{2020MNRAS.492.5234Z} and the method in \citet{2025arXiv250801384H, 2025ApJ...993...40X}, we estimated the intrinsic disk and Comptonized fluxes as
\begin{equation*}
    F_{\rm disk}=F_{\rm diskbb},
\end{equation*}
and
\begin{equation*}
    F_{\rm Comp}=F_{\rm thcomp}-(1-c_f)F_{\rm diskbb},
\end{equation*}
where $F_{\rm thcomp}$ denotes the flux of the convolved \texttt{thcomp}$\otimes$\texttt{diskbb} component and $c_f$ is the covering fraction in \texttt{thcomp}. The subtraction removes the direct, unscattered disk contribution included in the convolved spectrum.

Representative spectral fits for 4U~1543$-$47, illustrating the evolution of the disk and Comptonized components during the outburst, are presented in Appendix~\ref{sec:spec_timing_appendix}.

\subsection{Temporal Analysis}

To investigate the evolution of variability during the decay phase, we analyzed \textit{RXTE}/PCA event data from the selected observations. The event files were screened using the same GTIs generated in the standard data reduction procedure with \texttt{maketime}. We computed averaged power density spectra (PDS) over the full PCA energy band using \texttt{powspec} in \texttt{HEASoft} version 6.31\footnote{\url{https://heasarc.gsfc.nasa.gov/docs/software/lheasoft/}}. For the timing analysis, the light curves were divided into contiguous 32~s segments with a time resolution of 1/64~s, corresponding to a frequency range from 1/32~Hz to the Nyquist frequency of 32~Hz.

Because the archival sample contains a large number of observations and reliable multi-component Lorentzian modeling requires manual inspection, we adopted a targeted timing analysis. We restricted this analysis to the outbursts for which the disk-recession tracer can be measured, namely those with a well-sampled soft state exponential disk decay and sufficient coverage after the soft state. In this subset, nearly all outbursts with adequate post-soft-state coverage show a measurable departure of the disk luminosity from the fitted exponential baseline. Sources without soft state coverage, or outbursts lacking post-soft-state observations, were not included in the timing comparison because the exponential baseline and the subsequent deficit cannot be defined robustly.

For the quantitative timing analysis, the PDS were converted to the fractional-rms normalization \citep{1991ApJ...383..784M} after subtracting the Poisson noise contribution and were fitted with a sum of Lorentzian components. Following the methodology of \citet{2002ApJ...572..392B}, the broadband noise was described with zero-centered Lorentzians. We initially allowed up to three broadband noise components and removed components that were not required by the data. The F-test was used only as a practical model-simplification criterion for these broadband noise components. To avoid retaining spurious weak components, we further required the normalization significance of each retained broadband noise component to satisfy ${\rm Norm}/\sigma_{\rm Norm} > 1.5$.

Additional Lorentzian components were included to describe low-frequency QPOs and possible sub-harmonic features when narrow or peaked residuals were present in the PDS. QPO candidates were retained only when their fitted normalization satisfied ${\rm Norm}/\sigma_{\rm Norm} > 2.0$,
followed by manual inspection of the fitted PDS to reject obvious misidentifications or components driven by poorly constrained continuum noise.

For each Lorentzian component, we use the characteristic frequency
\begin{equation}
\nu_{\rm char} = \sqrt{\nu_0^2+\Delta^2},
\end{equation}
where $\nu_0$ is the centroid frequency and $\Delta$ is the half-width at half-maximum. This quantity is used as the timing frequency throughout the correlation analysis.

Representative spectral and timing fits are shown in Appendix~\ref{sec:spec_timing_appendix} for 4U~1543$-$47. For display purposes, the PDS in those illustrative panels is shown in the Leahy normalization, for which the nominal Poisson noise level is 2 \citep{1983ApJ...266..160L}. This choice visually clarifies the transition from Poisson-noise-dominated soft state PDS to intrinsically variable Compton-dominated PDS. The quantitative timing parameters used in Section~\ref{sec:variability_dis}, however, are derived from the fractional-rms-normalized fits described above.

\section{Analytical Method for Tracing Disk Recession}
\label{sec:analytical_method}

To trace the geometrical evolution of the accretion flow, we use the global energy budget of the thermal disk. In a geometrically thin accretion disk, the disk luminosity is primarily determined by the mass inflow rate and by the characteristic radius at which gravitational energy is released \citep{1973A&A....24..337S}. To a Newtonian approximation, the bolometric disk luminosity scales as
\begin{equation}
L_{\rm d} \propto \frac{G M_{\rm BH}\dot{M}}{R_{\rm tr}},
\label{eq:Ld_general}
\end{equation}
where $G$ is the gravitational constant, $M_{\rm BH}$ is the black hole mass, $\dot{M}$ is the mass accretion rate through the disk, and $R_{\rm tr}$ is the disk truncation radius. General relativistic effects, the inner boundary condition, and spectral hardening modify the absolute radiative efficiency, especially near the innermost stable circular orbit (ISCO) \citep{1995ApJ...445..780S, 2000MNRAS.313..193M}. Nevertheless, Equation~\ref{eq:Ld_general} captures the essential inverse dependence of the thermal disk luminosity on the characteristic inner radius.

During the soft state, the optically thick disk is generally expected to extend close to the ISCO, so that $R_{\rm tr}\simeq R_{\rm ISCO}$ remains approximately constant. In this regime, variations in disk luminosity primarily track the evolution of the mass accretion rate. The approximately exponential decay observed during the soft state can therefore be used as a fiducial baseline for the underlying mass inflow,
\begin{equation}
L_{\rm d}^{\rm exp}(t) = A \exp\left[-\frac{t-t_{\rm start}}{\tau}\right],
\label{eq:exp_baseline}
\end{equation}
where $A$ is the normalization, $\tau$ is the decay timescale, and $t_{\rm start}$ is the start time of the fitted exponential interval. It is also consistent with the expectation from irradiated disk decay models, in which X-ray irradiation can keep the outer disk in a hot, ionized state and allow the disk to drain viscously \citep{1998MNRAS.293L..42K}.

Following the approach of \citet{2023Sci...381..961Y}, we assume that this soft state exponential baseline approximately persists through the subsequent transition.
Combined with Equation~\ref{eq:Ld_general}, a thermal disk luminosity deficit relative to the extrapolated baseline can be interpreted as a reduction in the disk accretion efficiency due to the outward recession of the optically thick disk. The extrapolated luminosity $L_{\rm d}^{\rm exp}(t)$ represents the disk luminosity expected for the same fiducial mass inflow if the disk had remained at $R_{\rm ISCO}$. Comparing this expected luminosity with the observed disk luminosity $L_{\rm d}^{\rm obs}(t)$ gives
\begin{equation}
\frac{R_{\rm tr}(t)}{R_{\rm ISCO}}
=
\frac{L_{\rm d}^{\rm exp}(t)}{L_{\rm d}^{\rm obs}(t)} .
\label{eq:Rtr_tracer}
\end{equation}

The unabsorbed disk fluxes calculated in Section~\ref{sec:data} were converted to disk luminosities assuming the distances listed in Table~\ref{tab:sourcesinfo}. For sources lacking specific distance measurements, we adopted a typical distance of 8~kpc. Since the same distance factor enters both $L_{\rm d}^{\rm exp}$ and $L_{\rm d}^{\rm obs}$ for a given source, the inferred recession tracer is insensitive to the absolute distance, although the distance uncertainty affects luminosities normalized to the Eddington value.

For each selected outburst or decay episode, we first identified a soft state interval during which the disk luminosity declines smoothly and approximately exponentially. Because standard accretion-state classifications can depend on spectral models and band definitions, the exponential baseline interval was selected primarily from the disk light-curve morphology. Specifically, the selected interval is required to be disk-dominated, to show an approximately linear decline in $\log L_{\rm d}$, and to precede the final rapid thermal deficit associated with the soft-to-hard transition.

In sources with reflares or multiple smooth decay segments, we used the evolution of the fitted spectral parameters as an additional guide to identify the final exponential decay segment before the transition. In particular, abrupt changes in $T_{\rm in}$, the \texttt{diskbb} normalization, and the covering fraction $c_f$ in \texttt{thcomp} were used as reference points for the onset of the final transition. This avoids fitting earlier decay segments that are subsequently interrupted by reflares or renewed disk brightening, and ensures that the fitted baseline corresponds to the disk evolution immediately preceding the thermal luminosity deficit.

The start and end times of the exponential window, together with the decay parameters $A$ and $\tau$ in Equation~\ref{eq:exp_baseline}, were determined through the Markov Chain Monte Carlo (MCMC) fitting procedure described in Appendix~\ref{sec:fit_exp_baseline}. The end of the fitted exponential window, $t_{\rm end}$, marks the onset of the deficit, and the corresponding Eddington-scaled baseline luminosity defines $\dot{m}_{\rm break}$. The later disk luminosities were then compared with the extrapolated exponential baseline; their systematic deficit relative to this baseline is used to construct the model-defined recession tracer.

Some sources, such as XTE~J1817$-$330, XTE~J1859$+$226, GX~339$-$4, and GRO~J1655$-$40, show secondary flares or reflares superposed on the overall decay. These rebrightenings may be related to the complex viscous response of the irradiated outer disk, possibly including delayed mass inflow from outer disk regions or tidal effects at the disk edge \citep[e.g.,][]{1998MNRAS.293L..42K, 2002MNRAS.337.1329T, 2008MNRAS.388..753G}. For such cases, we avoided fitting intervals strongly affected by reflares and instead used the smooth decay segment that best represents the underlying exponential trend before the final thermal-deficit and Comptonized-component flare.

\section{Results} \label{sec:results}

\subsection{Decaying light curves and disk recession}
\label{subsec:decay_lc}

\begin{figure*}[htbp]
    \centering
    \includegraphics[width=0.95\textwidth]{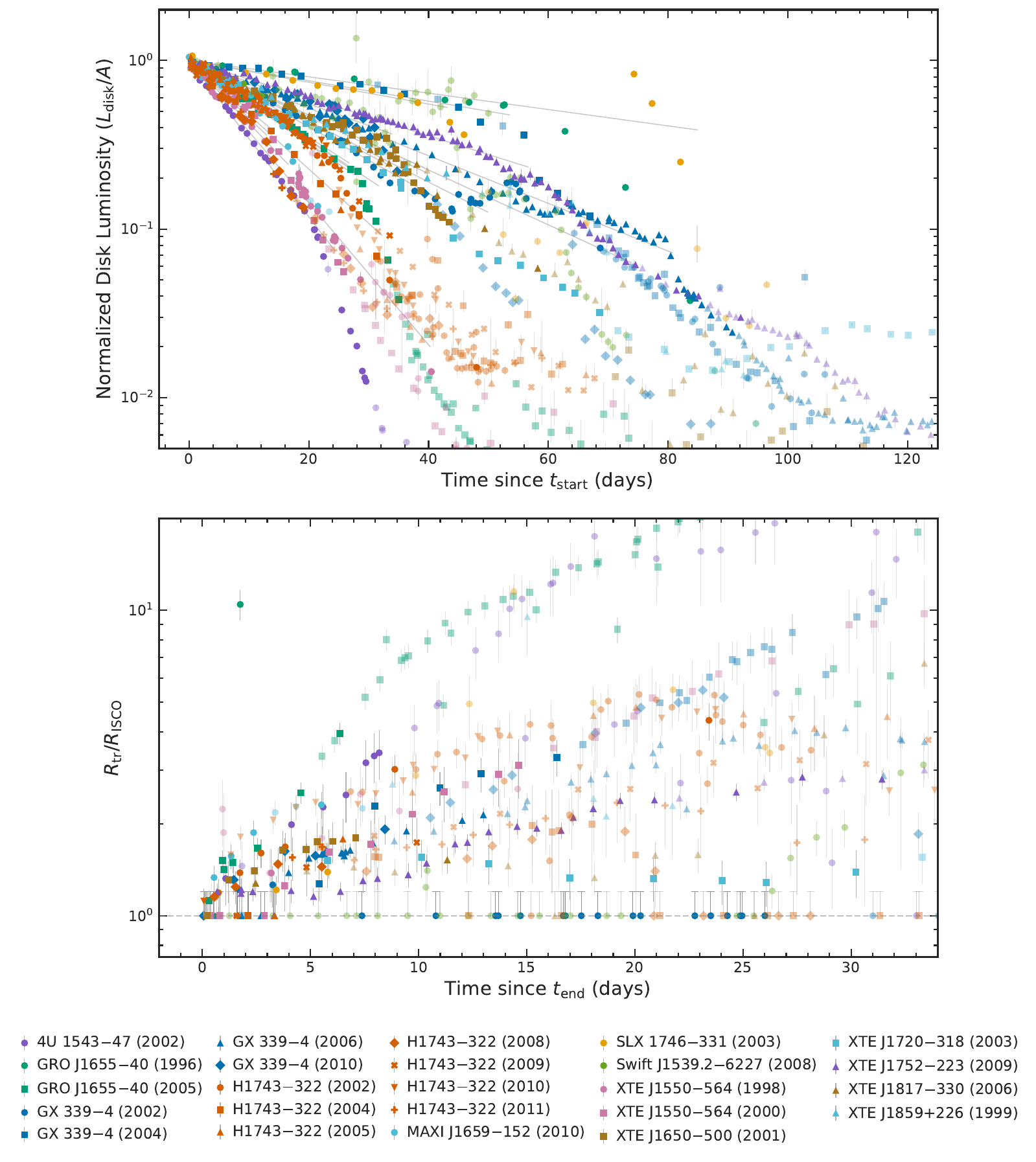}
    \caption{
    Normalized disk luminosity and inferred disk-recession evolution for the selected outbursts.
    \textbf{Panel A}: disk luminosity normalized by the best-fit exponential normalization, $L_{\rm d}/A$, as a function of time since $t_{\rm start}$, the beginning of the fitted soft state exponential decay interval. The faint solid curves show the best-fit exponential decay models for individual outbursts.
    \textbf{Panel B}: inferred truncation radius, $R_{\rm tr}/R_{\rm ISCO}$, as a function of time since $t_{\rm end}$, the end of the fitted exponential interval and the model-defined onset of the disk-luminosity deficit. The horizontal dashed line marks $R_{\rm tr}=R_{\rm ISCO}$. Points shown as lower limits correspond to epochs for which the inferred radius is smaller than or statistically consistent with the ISCO. Translucent markers indicate epochs for which the disk parameters are weakly constrained. Different outbursts from the same source are shown with the same color but with different markers.
    }
    \label{fig:disk_receding}
\end{figure*}

Following the procedure described in Section~\ref{sec:analytical_method}, we fitted the soft state disk decay of each selected outburst with an exponential function. The best-fit parameters are summarized in Appendix Table~\ref{tab:fit_results}. We retained only episodes with at least two valid disk-luminosity measurements after $t_{\rm end}$, so that the post-break deficit and the corresponding recession tracer can be evaluated. The fitted decay timescales span a broad range, from $\sim10$ to $\sim90$ days, indicating that the viscous decay properties are not universal across the sample. Nevertheless, after normalizing the disk luminosity by the best-fit normalization $A$ and shifting the time axis by $t_{\rm start}$, the outbursts show a common phenomenology: an approximately exponential soft state decay followed by a systematic deficit relative to the extrapolated baseline.

This behavior is summarized in Figure~\ref{fig:disk_receding}. Panel A shows the normalized disk luminosity, $L_{\rm d}/A$, together with the best-fit exponential baselines. In most cases, the disk luminosity initially follows an exponential trend and then drops below the extrapolated soft state baseline as the source evolves toward the hard state. This late-time thermal deficit is the primary observational feature used in this work to trace the recession of the optically thick disk.

Panel B shows the corresponding recession tracer, $R_{\rm tr}/R_{\rm ISCO}$, computed from Equation~\ref{eq:Rtr_tracer}, as a function of time since $t_{\rm end}$. Here $t_{\rm end}$ is the end of the fitted exponential interval and marks the onset of the disk-luminosity deficit in our fitting framework. When the observed disk luminosity is higher than, or statistically consistent with, the extrapolated baseline, the inferred radius is consistent with the ISCO and is plotted at the physical lower bound, $R_{\rm tr}=R_{\rm ISCO}$.

The individual light-curve fits for the selected outbursts are shown in Appendix~\ref{sec:individual}. These figures provide the detailed fit quality for each source, while Figure~\ref{fig:disk_receding} highlights the common evolution across the sample. The constrained post-$t_{\rm end}$ measurements of $R_{\rm tr}/R_{\rm ISCO}$ for all 24 outbursts are provided in Appendix Table~\ref{tab:rtr_values} (published in its entirety in machine-readable format).

\subsection{Evolution of Variability Properties}
\label{sec:variability_dis}

\begin{figure*}[!t]
    \centering
    \includegraphics[width=0.95\textwidth]{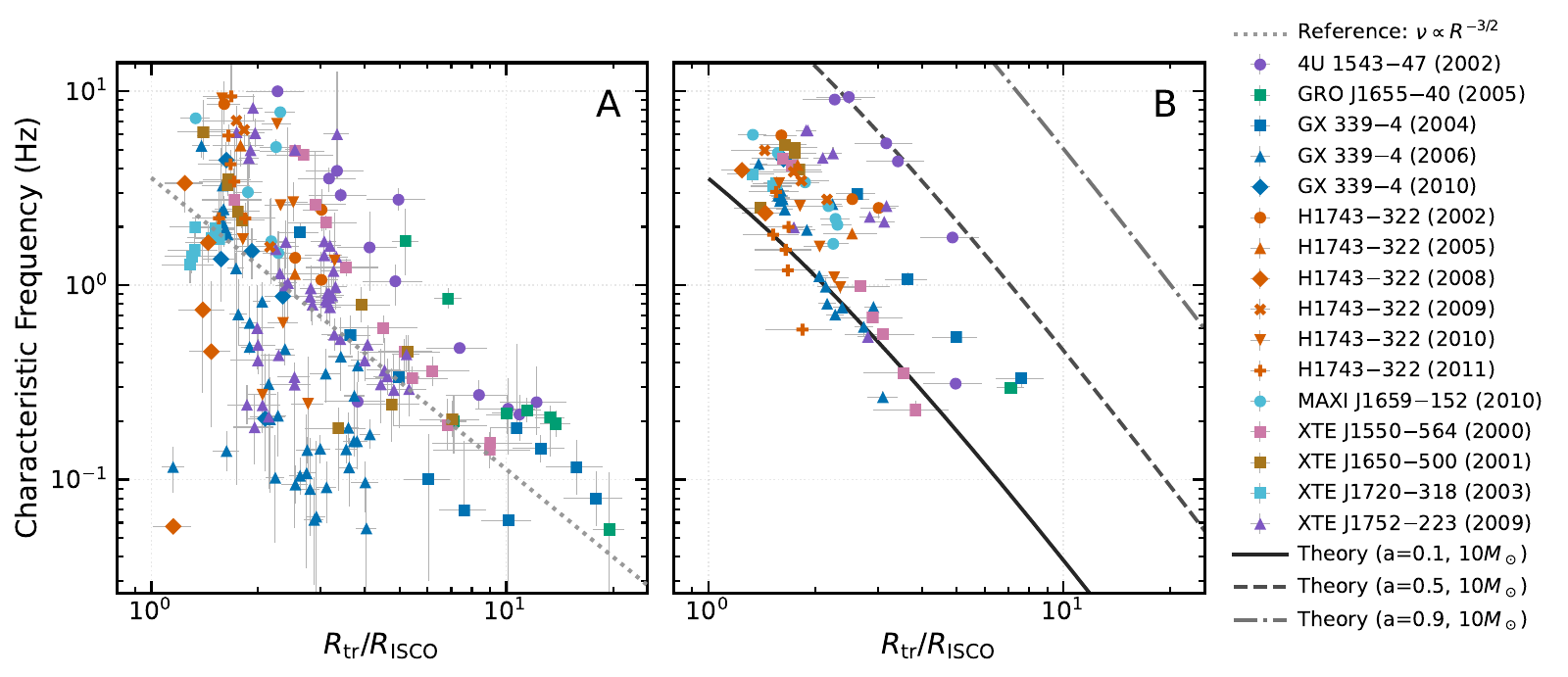}
    \caption{
    Timing characteristic frequencies as a function of the inferred disk truncation radius $R_{\rm tr}/R_{\rm ISCO}$.
    \textbf{Panel A}: characteristic frequency of the broadband-noise component. The dashed line shows a $\nu\propto R^{-3/2}$ scaling expected for a Keplerian dynamical frequency, plotted only as a visual reference.
    \textbf{Panel B}: QPO characteristic frequency. The grey curves show the expected frequencies from the rigid Lense--Thirring precession model for representative black hole spins \citep{2009MNRAS.397L.101I}.
    }
    \label{fig:rtr-timing}
\end{figure*}

We next compare the disk truncation radius inferred from the luminosity deficit with the evolution of the timing properties. Figure~\ref{fig:rtr-timing} shows the characteristic frequencies of the broadband noise and QPO components as a function of $R_{\rm tr}/R_{\rm ISCO}$.

As a visual reference, the representative spectra and PDS shown in Appendix~\ref{sec:spec_timing_appendix} illustrate the spectral--timing evolution during the transition. In the disk-dominated soft state, the PDS are largely consistent with counting noise, whereas broadband noise and low-frequency QPO features become prominent when the Comptonized component is enhanced. As the source continues to evolve toward the hard state, these timing features shift to lower frequencies. This behavior is consistent with the general trend shown in Figure~\ref{fig:rtr-timing}, where the characteristic frequencies decrease as the inferred disk truncation radius increases.

The broadband noise frequencies generally decrease as the inferred disk truncation radius increases, indicating that the dominant variability timescale becomes longer during the disk recession phase. As a reference, Panel A of Figure~\ref{fig:rtr-timing} also shows the expected $\nu \propto R^{-3/2}$ scaling for a Keplerian dynamical frequency. The observed broadband noise frequencies are broadly consistent with a decreasing trend, although they exhibit substantial scatter and should not be interpreted as direct measurements of the local Keplerian frequency. In particular, some low-frequency components may be affected by red-noise leakage or by the decomposition of broad continuum variability. A similar decreasing trend is observed for the QPO frequencies when they can be measured. Although the number of simultaneous or near-simultaneous spectral--timing measurements is limited, this behavior is qualitatively consistent with a characteristic variability radius moving outward as the thin disk recedes.

For qualitative reference, Panel B of Figure~\ref{fig:rtr-timing} also shows the QPO frequencies predicted by the Lense--Thirring precession model of a hot inner flow \citep{2009MNRAS.397L.101I} for several representative black hole spins. Under the adopted assumptions for the hot-flow geometry, the observed QPO frequencies lie close to the predicted tracks for relatively low spins, $a_*\sim0.1$--$0.5$. This comparison is intended only as a qualitative reference, since the predicted frequencies depend on the assumed mass, spin, surface-density profile, and radial extent of the precessing flow. We note that relatively low black-hole spins have also been discussed recently in the broader context of the tension between spin measurements in X-ray binaries and gravitational-wave sources \citep{2026NewAR.10201746Z}.

Taken together, the spectral and timing evolution shows a coherent qualitative trend. The thermal disk luminosity drops below the exponential soft state baseline, the Comptonized component is enhanced, and the characteristic variability frequencies shift to lower values. These trends are consistent with a soft-to-hard transition accompanied by the recession of the optically thick disk and the growth of a hot inner accretion flow.

\section{Discussion} \label{sec:discussion}

The thermal disk luminosity deficits identified in this work provide a means to examine the evolution of accretion geometry during the soft-to-hard transition. In our framework, the soft state exponential decay defines a fiducial baseline for the underlying mass inflow, and the resulting thermal luminosity deficit is converted into a model-defined tracer of disk recession. The accompanying enhancement of the Comptonized component and the decrease of characteristic timing frequencies are qualitatively consistent with a scenario in which the optically thick disk recedes while the hot inner flow expands. In the following, we discuss the diversity of the inferred recession behavior, compare our tracer with direct spectral inner-radius constraints, and examine the roles of outer-disk cooling and disk radiative efficiency.

\subsection{Diversity in the Onset and Evolution of Disk Recession}

The inferred disk recession behavior shows substantial diversity across sources and even among repeated outbursts of the same source, especially for GX~339$-$4. We use two quantities to characterize this diversity. The first is the break accretion rate, $\dot{m}_{\rm break}$, defined as the Eddington-scaled baseline luminosity at the end of the fitted exponential-decay interval:
\begin{equation}
    \dot{m}_{\rm break} \equiv \dot{m}_{\rm d}^{\rm exp}(t_{\rm end}),
\end{equation}
where $\dot{m}_{\rm d}^{\rm exp}=\dot{M}_{\rm d}^{\rm exp}/\dot{M}_{\rm Edd}=L_{\rm d}^{\rm exp}/L_{\rm Edd}$ and $\dot{M}_{\rm Edd}=L_{\rm Edd}/(\eta c^2)$ with a reference efficiency $\eta=0.1$. The second quantity is the recession slope $\gamma$, which describes how rapidly the inferred truncation radius changes as the expected exponential disk accretion rate declines. We parameterize this evolution as
\begin{equation}
    \frac{R_{\rm tr}}{R_{\rm ISCO}} = \left( \frac{\dot{m}_{\rm d}^{\rm exp}}{\dot{m}_{\rm break}} \right)^{-\gamma}.
    \label{eq:rtr_mdot_fit}
\end{equation}
A larger value of $\gamma$ therefore indicates a faster increase of $R_{\rm tr}/R_{\rm ISCO}$ for a given decrease in $\dot{m}_{\rm d}^{\rm exp}$ after the onset of the deviation. Equivalently, because $L_{\rm d}^{\rm obs}/L_{\rm d}^{\rm exp} = R_{\rm ISCO}/R_{\rm tr}$, Equation~\ref{eq:rtr_mdot_fit} describes a power-law decline in the effective radiative efficiency of the thin disk below $\dot{m}_{\rm break}$, $\eta_{\rm disk} \propto (\dot{m}_{\rm d}^{\rm exp}/\dot{m}_{\rm break})^{\gamma}$, taking the same functional form as the efficiency scaling for radiatively inefficient accretion adopted in previous studies \citep[e.g.,][]{2011MNRAS.414..677C, 2012MNRAS.427.1580X, 2014MNRAS.437.3087K, 2015MNRAS.450.3410E}; we discuss this connection in detail in Section~\ref{subsec:rad-eff}.

The fit was performed in logarithmic space using post-$t_{\rm end}$ data points with $R_{\rm tr}/R_{\rm ISCO}>1.05$, with the model anchored at $R_{\rm tr}=R_{\rm ISCO}$ when $\dot{m}_{\rm d}^{\rm exp}=\dot{m}_{\rm break}$. Uncertainties were estimated by Monte Carlo resampling of the exponential baseline and disk luminosities. For each realization, we drew one set of decay parameters from the MCMC posterior, perturbed the observed disk luminosities according to their asymmetric errors, recomputed $R_{\rm tr}/R_{\rm ISCO}$ and $\dot{m}_{\rm d}^{\rm exp}/\dot{m}_{\rm break}$, and refitted the slope with inverse-variance weights in log space. Weakly constrained disk points were kept but assigned twice-as-large logarithmic uncertainties. The median and 16th--84th percentiles of the resulting $\gamma$ distribution were adopted as the best value and $1\sigma$ uncertainty.

\begin{figure*}[!htbp]
\centering
\includegraphics[width=0.95\textwidth]{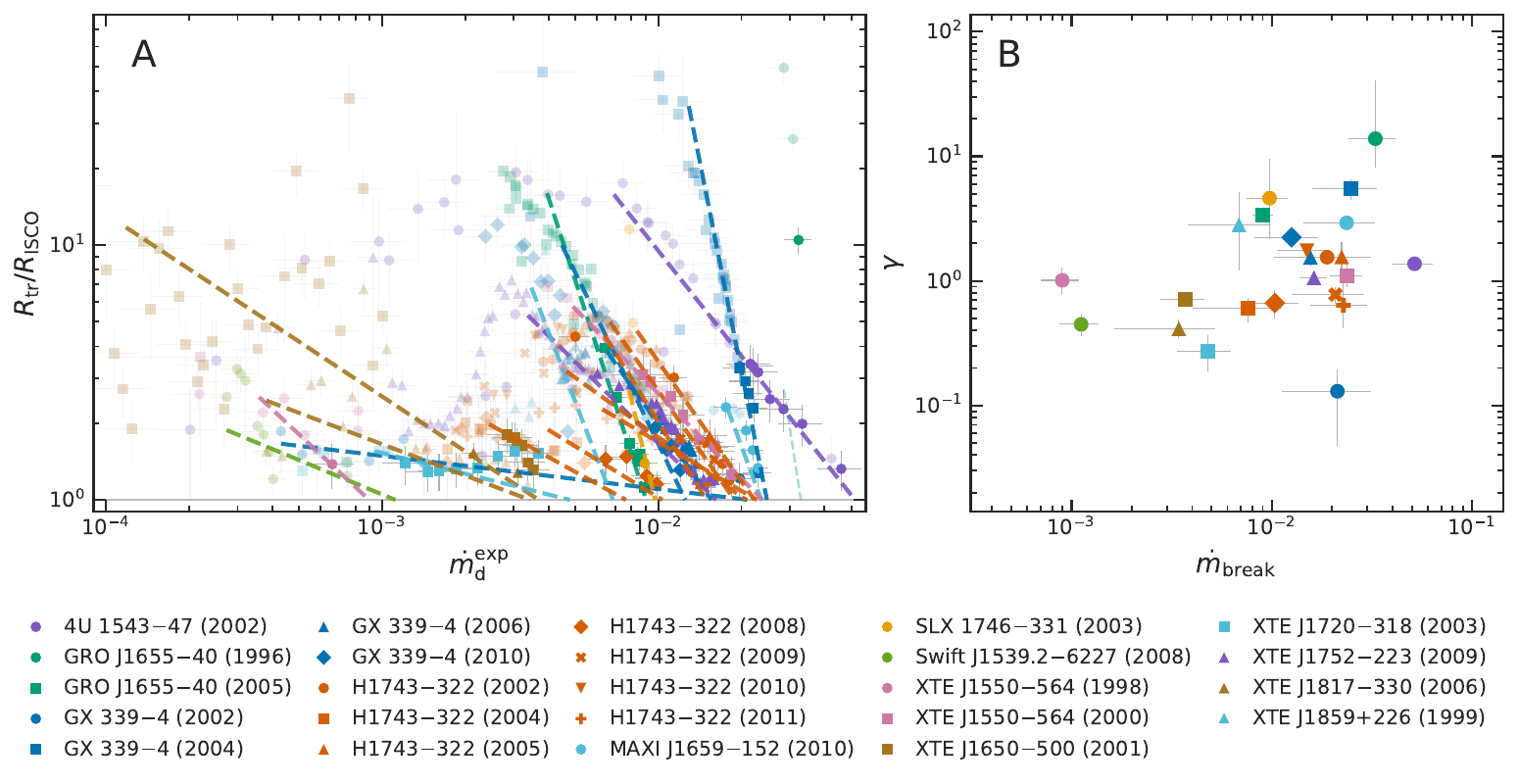}
\caption{Diversity in the inferred disk recession evolution. \textbf{Panel A}: inferred recession tracer, $R_{\rm tr}/R_{\rm ISCO}$, as a function of the fiducial Eddington-scaled baseline, $\dot{m}_{\rm d}^{\rm exp}$. Dashed lines show the posterior-median recession models for individual outbursts. \textbf{Panel B}: recession slope, $\gamma$, as a function of the break scale, $\dot{m}_{\rm break}$.}
\label{fig:rtr-mdot}
\end{figure*}

Figure~\ref{fig:rtr-mdot} summarizes the inferred recession evolution and compares the recession slope $\gamma$ with the break scale $\dot{m}_{\rm break}$. The values of $\dot{m}_{\rm break}$ span a broad range, from $\sim10^{-3}$ to $\sim10^{-1}$. This broad distribution suggests that the onset of the thermal luminosity deficit is not set by a single universal Eddington-scaled luminosity. Although hot-flow formation is often discussed in the context of advection-dominated accretion flow (ADAF) solutions \citep{1995ApJ...452..710N, 2014ARA&A..52..529Y}, we use $\dot{m}_{\rm break}$ here only as an empirical scale marking the departure of the thermal disk emission from the soft state exponential decay. Its range is broadly comparable to the luminosity interval over which soft-to-hard transitions are observed in BHXRBs \citep[e.g.,][]{2003A&A...409..697M, 2006MNRAS.370..837G}.

The diagram also allows a comparison among repeated outbursts from the same source. The seven outbursts of H1743$-$322 occupy a relatively compact region in the $\gamma$--$\dot{m}_{\rm break}$ plane, whereas the four outbursts of GX~339$-$4 and the two outbursts each of GRO~J1655$-$40 and XTE~J1550$-$564 are more widely distributed. This suggests that the degree of intra-source diversity differs from source to source, although the present sample is too small to separate source-dependent effects from outburst-to-outburst variations.

Although $\dot{m}_{\rm break}$ provides a useful scale for comparing different outbursts, the fitted recession slope $\gamma$ shows no significant correlation with $\dot{m}_{\rm break}$ in the present sample. We quantified this using the Spearman rank correlation coefficient with Monte Carlo propagation of the measurement uncertainties \citep{2014arXiv1411.3816C}. The resulting coefficient is $r_{\rm S}=0.3$, with a significance of only $0.6~\sigma$. Thus, the rate at which the inferred truncation radius grows after the onset of the deficit does not appear to be controlled solely by the accretion-rate scale at which the deficit begins.

We note, however, that $\dot{m}_{\rm break}$ and $\gamma$ are not statistically independent quantities, because both are derived from the same exponential baseline. The comparison between $\dot{m}_{\rm break}$ and $\gamma$ should therefore be regarded as an exploratory diagnostic of the diversity among outbursts, rather than as a test of a direct relation. The absence of a significant correlation suggests that additional source- and outburst-dependent factors are important. These may include the irradiation geometry, the size and thermal state of the outer disk, the efficiency of disk evaporation, the magnetic field configuration, and the detailed accretion history preceding the soft-to-hard transition.

\subsection{Spectral Constraints on the Inner Disk Radius}
\label{subsec:spectral-radius}

In principle, the inner disk radius can be estimated directly from the \texttt{diskbb} normalization \citep{1984PASJ...36..741M, 1998PASJ...50..667K}:
\begin{equation}
   R_{\rm in} = \kappa^2 \xi \sqrt{N_{\rm diskbb} D_{10}^{2} / \cos i},
   \label{eq:rin}
\end{equation}
where $\kappa$ is the spectral hardening factor, $\xi$ is the inner-boundary correction factor accounting for the zero-torque inner boundary condition, $D_{10}$ is the distance in units of $10\,\mathrm{kpc}$, and $i$ is the inclination. For a thin disk extending down to the ISCO, \citet{1998PASJ...50..667K} estimated $\xi \approx 0.41$, which is approximately valid for zero black hole spin but becomes inaccurate at high spins \citep{2026NewAR.10201746Z}. When the thin disk is truncated and transitions into an inner hot flow, however, the viscous torque is not expected to vanish at the truncation radius (the zero-stress condition applies only near the ISCO), so $\xi \sim 1$ \citep{2016MNRAS.458.2199B}. In the following discussion of truncated disks during the transition, we adopt $\xi \approx 1$. Converting the apparent radius into a physical radius therefore requires assumptions about $\kappa$ and $\xi$, which become particularly uncertain during the soft-to-hard transition when the thermal disk weakens and the Comptonized component dominates the 3--25\,keV PCA band.

As illustrated in Figure~\ref{fig:4u1543_paras}, the spectral evolution of 4U~1543$-$47 during its 2002 outburst highlights this limitation. Immediately after the end of the soft state exponential decay ($t_{\rm end}$), the fitted \texttt{diskbb} normalization drops substantially while $T_{\rm in}$ shows an apparent increase. Simultaneously, the covering fraction $c_f$ rises rapidly from $<0.1$ to $\sim 0.6$--$1.0$, indicating that a larger fraction of the seed photons from the disk is scattered by the developing hot inner flow. At later epochs, the thermal component becomes weakly constrained, and both $T_{\rm in}$ and $N_{\rm diskbb}$ exhibit large uncertainties and scatter. If interpreted literally under a constant $\kappa$ and $\xi$, the drop in $N_{\rm diskbb}$ after $t_{\rm end}$ would imply an inward contraction of the disk, in apparent contradiction with the disk recession inferred from the luminosity deficit.

To test the reliability of direct spectral radius estimates across the sample and compare them with our luminosity-deficit tracer, we constructed a relative spectral inner radius, $R_{\rm in,spec} \equiv \sqrt{N_{\rm diskbb}/N_{\rm diskbb,soft}}$, where $N_{\rm diskbb,soft}$ is the median of the well-constrained \texttt{diskbb} normalizations within each outburst's soft state exponential decay window. With constant $\kappa$ and $\xi$, $R_{\rm in,spec}$ would trace $R_{\rm in}/R_{\rm ISCO}$. As an extension of Figure~\ref{fig:rtr-timing}, Figure~\ref{fig:cosistency_check} compares both the broadband-noise frequency ($\nu_{\rm LFN}$; left two columns) and the QPO frequency ($\nu_{\rm QPO}$; right two columns) against the luminosity-derived $R_{\rm tr}/R_{\rm ISCO}$ and against $R_{\rm in,spec}$ for three fixed coronal electron temperatures ($kT_{\rm e} = 50$, $150$, and $300\,\mathrm{keV}$).

In contrast to the monotonic negative correlation obtained with $R_{\rm tr}/R_{\rm ISCO}$, $R_{\rm in,spec}$ exhibits large scatter and a strong $kT_{\rm e}$ dependence. Many transition observations yield $R_{\rm in,spec} < 1$, particularly at lower $kT_{\rm e}$. Interpreting such values literally as physical disk contraction would require the inner disk edge to move inward during the state transition and require truncation during the soft state. Moreover, several outbursts exhibit an apparent positive trend between the characteristic timing frequencies ($\nu_{\rm LFN}$ and $\nu_{\rm QPO}$) and $R_{\rm in,spec}$, opposite to the expected timescale--geometry relation. This behavior reflects a strong degeneracy between $N_{\rm diskbb}$ and $T_{\rm in}$ when only the high-energy tail of a weakened, strongly Comptonized disk is sampled above $3\,\mathrm{keV}$.

Similar departures from the standard $L_{\rm d} \propto T_{\rm in}^4$ scaling when the thermal component is no longer dominant have been widely reported in previous \textit{RXTE} studies \citep[e.g.,][]{2000MNRAS.313..193M, 2011MNRAS.411..337D} and are commonly attributed to an increase in the spectral hardening factor $\kappa$ as the corona strengthens, together with parameter degeneracies in the spectral decomposition. From Equation~\ref{eq:rin} and the definition of the spectral hardening factor ($T_{\rm in} \propto \kappa T_{\rm eff}$; \citealt{1995ApJ...445..780S}, \citealt{2019ApJ...874...23D}), one has $N_{\rm diskbb} \propto R_{\rm in}^2 \kappa^{-4}$ for a fixed $\xi$. Thus, an increase in $\kappa$ simultaneously raises the apparent $T_{\rm in}$ and sharply reduces $N_{\rm diskbb}$ even if the physical inner radius does not shrink.
In contrast, the bolometric disk luminosity $L_{\rm d} \propto N_{\rm diskbb} T_{\rm in}^4$ does not depend explicitly on $\kappa$, remains much more stable across $kT_{\rm e}=50$--$300\,\mathrm{keV}$, and preserves a monotonic anti-correlation with independent timing frequencies (Figure~\ref{fig:cosistency_check}), making it a more suitable tracer of disk geometry when individual spectral parameters cannot be reliably constrained.

\subsection{Outer Disk Cooling and Inner Disk Recession}

In irradiated disk models, the outer disk can remain in the hot, ionized state as long as irradiation from the central X-ray source is sufficiently strong. Under this condition, the disk drains viscously, and the central accretion rate can approximately follow an exponential decay \citep{1998MNRAS.293L..42K}. When the irradiation becomes too weak to keep the outer disk hot, the hot region shrinks, and the accretion rate declines more rapidly, approaching a linear decline in simplified irradiated-disk models.

However, the deviation observed in our sample is not simply a decline in the total X-ray output. In most outbursts, the thermal disk luminosity drops rapidly while the Comptonized component becomes enhanced. This behavior suggests a redistribution of the accretion power from the optically thick disk to a hot inner flow, rather than a uniform weakening of the central irradiation. Moreover, the observed flux depends on the radiative efficiencies and energy partition between the thin disk and the inner hot flow, which are expected to change during the state transition.

To test whether the disk luminosity deviation can be explained solely by the loss of irradiation support in the outer disk, we estimate an outer-disk irradiation scale at the onset of the deviation, following the irradiated disk instability model (DIM) framework \citep{1982ApJ...260L..83C, 1999MNRAS.303..139D, 2001A&A...373..251D, 2001NewAR..45..449L}.
For an irradiated disk, the surface temperature can be written as
\begin{equation}
    T_{\rm sur}^4 = T_{\rm eff}^4 + T_{\rm irr}^4 .
\end{equation}
The irradiation temperature may be expressed as
\begin{equation}
    T_{\rm irr}^4(R) =
    \frac{L_{\rm X}(1-\beta)}{4\pi\sigma R^2}
    \left(\frac{H}{R}\right)^2
    \left(\frac{{\rm d}\ln H}{{\rm d}\ln R}-1\right),
\end{equation}
where \(\beta\) is the disk albedo, \(H/R\) describes the disk geometry, and \(\sigma\) is the Stefan--Boltzmann constant. Because these geometric factors are uncertain, we adopt the commonly used parametrization
\begin{equation}
    T_{\rm irr}^4(R) =
    \frac{C_{\rm irr} L_{\rm X}}{4\pi\sigma R^2},
\end{equation}
where \(C_{\rm irr}\) is an effective outer-disk parameter that accounts for the effects of the illuminating spectrum, disk albedo, disk flaring, and irradiation geometry. We adopt \(C_{\rm irr}=5\times10^{-3}\), a typical value empirically inferred from BHXRB outburst light curves \citep[e.g.,][]{2018Natur.554...69T, 2018MNRAS.480....2T}, and define \(R_{\rm irr}\) as the radius at which \(T_{\rm irr}=10^4\,{\rm K}\).

\begin{figure}[ht]
\centering
\includegraphics[width=0.46\textwidth]{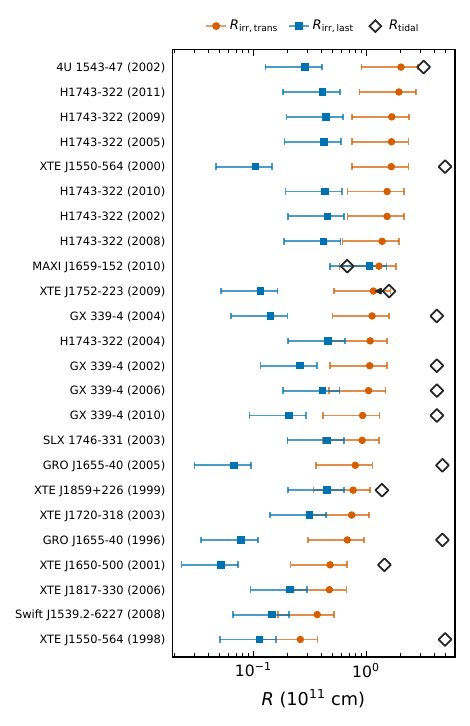}
\caption{
Comparison of irradiation radii at the start of the disk deviation, $R_{\rm irr,trans}$ (orange circles), and at the last observation, $R_{\rm irr,last}$ (blue squares). The irradiation radius is defined by $T_{\rm irr}=10^4$~K for a fiducial value of $C_{\rm irr}=5\times10^{-3}$. The horizontal bars indicate the corresponding ranges when varying $C_{\rm irr}$ from $10^{-3}$ to $10^{-2}$. Open black diamonds show the estimated outer-disk tidal truncation radius, $R_{\rm tidal} \approx 0.9\,R_{\rm L1}$, for systems with orbital period measurements (leftward arrow indicates an upper limit for XTE~J1752$-$223).
}
\label{fig:irradiation_radii}
\end{figure}

Figure~\ref{fig:irradiation_radii} presents the resulting irradiation radii. Using the total unabsorbed 0.01--1000\,keV model flux (\texttt{thcomp}$\otimes$\texttt{diskbb + gaussian}) interpolated in log space to $t_{\rm end}$, we find the irradiation radius at the onset of the deviation, $R_{\rm irr,trans} \sim 2.6\times10^{10}$--$2.0\times10^{11}\,\mathrm{cm}$ (typically $\sim 10^{11}\,\mathrm{cm}$) for our fiducial $C_{\rm irr}=5\times10^{-3}$; the horizontal bars in Figure~\ref{fig:irradiation_radii} show the corresponding range obtained by varying $C_{\rm irr}$ across the typical DIM interval of $10^{-3}$--$10^{-2}$ \citep{1999MNRAS.303..139D, 2001A&A...373..251D, 2018MNRAS.480....2T}. For the eight systems (13 outbursts) with orbital period constraints in Table~\ref{tab:sourcesinfo}, we compare $R_{\rm irr,trans}$ with the outer-disk tidal truncation radius $R_{\rm tidal} \approx 0.9\,R_{\rm L1}$ \citep{2002apa..book.....F}. In most systems, $R_{\rm irr,trans}$ lies below or comparable to $R_{\rm tidal}$, indicating that central irradiation is generally insufficient to keep the entire outer disk hot near $t_{\rm end}$ (except in the short-period system MAXI~J1659$-$152, where $R_{\rm irr,trans} > R_{\rm tidal}$). Thus, the irradiation radius comparison alone cannot rule out outer-disk cooling in most systems.

Beyond the overall uncertainty in the fiducial value of $C_{\rm irr}$, treating $C_{\rm irr}$ as a constant is itself a strong simplification \citep{1999MNRAS.303..139D, 2001A&A...373..251D, 2001NewAR..45..449L}. Relativistic X-ray reflection modeling of BHXRBs with broadened Fe~K$\alpha$ lines often reveals steep inner-disk emissivity profiles peaking near the ISCO due to strong general-relativistic light bending and compact corona--disk geometries, implying that the local irradiation efficiency is radially dependent. Applying a single constant $C_{\rm irr}$ to reconstruct the actual radial temperature profile of the disk would therefore produce severe errors, especially in the inner relativistic region. Even at large radii, outer-disk irradiation is physically complex and unlikely to remain constant across spectral states: multi-wavelength optical, near-IR, and UV observations of black hole and neutron-star binaries reveal significant state-dependent changes in the inferred irradiation efficiency and departures from simple constant-\(C_{\rm irr}\) prescriptions \citep{2008ApJ...688..537M, 2009MNRAS.392.1106G, 2019MNRAS.482..626K}.

A more natural interpretation is that outer-disk cooling and inner-disk evaporation are not mutually exclusive. Numerical calculations including both irradiation and evaporation show that disk truncation can strongly modify the late-time decay behavior \citep{2001A&A...373..251D}. Consistent with this view, \citet{2018MNRAS.480....2T} found that while the soft-state exponential decay is a robust feature of a fully irradiated viscous disk, the late-time break in a substantial fraction of BHXRB light curves cannot be explained by a standard constant-$C_{\rm irr}$ DIM cooling front alone and may instead reflect inner-disk evaporation into a radiatively inefficient flow or other processes in the inner X-ray emitting region. While outer-disk cooling likely contributes to the mass-transfer decay in systems where $R_{\rm irr,trans} \lesssim R_{\rm tidal}$, pure outer-disk cooling alone cannot explain the full behavior observed across the sample: (i) a sharp thermal deficit still occurs in short-period systems such as MAXI~J1659$-$152 where $R_{\rm irr,trans} > R_{\rm tidal}$; (ii) the drop in thermal disk luminosity is accompanied by an enhancement of the Comptonized component, pointing to a redistribution of accretion power from the optically thick disk to a hot inner flow rather than a uniform decline in total X-ray output; and (iii) the characteristic broadband-noise and QPO frequencies decrease systematically as the deficit grows (Figure~\ref{fig:rtr-timing}), consistent with an expanding hot inner flow. In this picture, the observed disk-luminosity deficit reflects the reduced effective accretion efficiency of the optically thick disk as its inner edge recedes, while outer-disk cooling and inner-disk evaporation can operate together during the soft-to-hard transition.

\subsection{Radiative Efficiency and Disk Recession}
\label{subsec:rad-eff}

From a more general perspective, the phenomenon we observed, a disk luminosity deficit during the state transition, corresponds to a reduction in the effective radiative efficiency of the thin disk. In a BHXRB, the accretion luminosity is commonly written as $L = \eta \dot{M} c^2$, where $\eta \sim 0.1$ for a thin disk extending down to the ISCO \citep{1973A&A....24..337S}. In the truncated disk plus inner ADAF scenario \citep{1994ApJ...428L..13N, 1997ApJ...489..865E}, the thermal luminosity is powered by the gravitational potential energy released down to the truncation radius $R_{\rm tr}$,
\begin{equation}
L_{\rm d} = \eta_{\rm disk}\dot{M}c^2 \propto \frac{G M_{\rm BH}\dot{M}}{R_{\rm tr}},
\label{eq:Ld_rad_eff}
\end{equation}
so that $\eta_{\rm disk}(R_{\rm tr}) \propto R_{\rm g}/R_{\rm tr}$ (where $R_{\rm g} = G M_{\rm BH}/c^2$ is the gravitational radius). A closely related analytical approach was used by \citet{2014MNRAS.437.3087K} to explain the apparent lack of short-period ($P_{\rm orb} < 4$\,h) BHXRBs relative to neutron-star systems. Starting from the same \citet{1998MNRAS.293L..42K} exponential decay for $\dot{M}(t)$, \citet{2014MNRAS.437.3087K} parameterized the switch to radiatively inefficient accretion below $f L_{\rm Edd}$ as either an instantaneous drop to $\eta = 0$ or a power-law decline ($\eta \propto \dot{M}^n$), showing that hiding short-period BHXRBs requires a steep efficiency drop ($n \gtrsim 3$) or abrupt spectral changes that shift emission out of the observing band. While both studies link a steepening luminosity decline to decreasing accretion efficiency, their scope and observables differ: \citet{2014MNRAS.437.3087K} modeled the total radiative efficiency to evaluate population-level detectability across $P_{\rm orb}$ without addressing $R_{\rm tr}$ directly, whereas we isolate the thermal disk component in individual outbursts to empirically trace $\eta_{\rm disk}(R_{\rm tr})$. Combining Equations~\ref{eq:rtr_mdot_fit} and \ref{eq:Ld_rad_eff} yields $\eta_{\rm disk}/\eta_{\rm ISCO} = R_{\rm ISCO}/R_{\rm tr} = (\dot{m}_{\rm d}^{\rm exp}/\dot{m}_{\rm break})^{\gamma}$, so our recession slope $\gamma$ directly plays the role of the power-law index $n$ in \citet{2014MNRAS.437.3087K}, though for the thermal disk component alone. Across the 23 well-constrained outbursts in Figure~\ref{fig:rtr-mdot}, $\gamma$ spans $0.13$--$5.46$ with a median of $\approx 1.1$ (16th--84th percentiles: $0.53$--$2.87$), broadly comparable to the power-law efficiency scalings inferred for hot accretion flows \citep[e.g., $n \sim 0.5$--$1.3$;][]{2012MNRAS.427.1580X}, hard-state radio--X-ray correlations \citep{2011MNRAS.414..677C}, and the factor of $\sim 2$ shortening of \textit{RXTE} decay timescales across the soft-to-hard transition \citep[corresponding to $\eta \propto \dot{M}^{\beta}$ with $\beta \sim 1$;][]{2015MNRAS.450.3410E}, while a subset of outbursts (e.g., GRO~J1655$-$40 2005, GX~339$-$4 2004, SLX~1746$-$331 2003, and the short-period system MAXI~J1659$-$152 with $\gamma \approx 2.9$) approach or exceed the steep efficiency drop ($n \gtrsim 3$) required by \citet{2014MNRAS.437.3087K}. Note, however, that because enhanced Comptonization partially compensates for the thermal disk decline during the transition, the total radiative efficiency typically drops less steeply than $\eta_{\rm disk}$. In this context, the broad scatter in our empirical onset scale $\dot{m}_{\rm break}$ ($\sim 10^{-3}$--$10^{-1}$) and recession slope $\gamma$ indicates that, although rapid declines in thermal efficiency are common during the soft-to-hard transition, both the onset and the rate of the drop vary substantially across sources and outbursts rather than following a single universal scaling.

Alternatively, $\eta_{\rm disk}$ can decrease without a large geometrical recession if the disk remains near the ISCO while an increasing fraction of the accretion power is dissipated in a corona or extracted by an outflow or jet \citep[e.g.,][]{1994ApJ...436..599S, 1999ApJ...510L.123B, 2001MNRAS.326..417M, 2002MNRAS.332..165M}. In an outflowing-corona geometry, mildly relativistic bulk motion away from the disk further beams coronal X-rays outward, suppressing downward irradiation and passive-disk reheating \citep{1999ApJ...510L.123B, 2001MNRAS.326..417M}. MAXI~J1820$+$070 provides a relevant example in the bright hard state, where reverberation lags and reflection evolution have been interpreted in terms of coronal geometry changes and outward bulk acceleration rather than a large change in inner disk radius \citep{2019Natur.565..198K, 2021NatCo..12.1025Y}. Energy extraction by an active or outflowing corona may therefore also contribute to the observed thermal deficit during the soft-to-hard transition. Nevertheless, such non-receding scenarios alone do not easily explain the simultaneous decrease in broadband-noise and QPO frequencies tracked alongside the luminosity deficit (Figure~\ref{fig:rtr-timing}), for which disk recession and an expanding hot inner flow remain a more unified interpretation.

\section{Conclusion}
\label{sec:conclusion}

The soft-to-hard state transition in BHXRBs is commonly associated with a change in the structure of the accretion flow, in which the optically thick disk recedes, and a hot inner flow develops. However, tracing this evolution directly from spectral fits is challenging because the thermal disk component weakens and becomes model-dependent near the hard state. In this work, we constructed a model-defined tracer of disk recession using the thermal disk luminosity deficit following the exponential decay during the soft state.

We inspected 26 BHXRBs observed with \textit{RXTE} and identified 24 outbursts with a well-defined soft state exponential baseline followed by a measurable decline in the thermal disk luminosity below the extrapolated trend. Adopting the fiducial assumption that the underlying disk mass inflow continues to follow the soft state exponential decay, we interpret the observed disk-luminosity deficit as an effective reduction in the thermal-disk accretion efficiency, which we parameterize as the outward recession of the optically thick disk. Under this framework, the evolution of the characteristic truncation radius during the soft-to-hard transition can be estimated through
\begin{equation*}
\frac{R_{\rm tr}(t)}{R_{\rm ISCO}}
=
\frac{L_{\rm d}^{\rm exp}(t)}{L_{\rm d}^{\rm obs}(t)} .
\end{equation*}
Although this inferred radius cannot be interpreted as a direct measurement of the disk inner edge, it provides a consistent empirical tracer against which the transition behavior can be compared across different outbursts.

The main results of this work are summarized as follows:
\begin{enumerate}
    \item The thermal disk luminosity initially follows an approximately exponential decay during the soft state and subsequently drops below the extrapolated baseline as the source evolves toward the hard state. This decline is frequently accompanied by an enhancement of the Comptonized component, consistent with a redistribution of the released accretion power from the optically thick disk toward a hot inner flow.

    \item In our framework, the inferred disk truncation radius increases rapidly after the thermal luminosity begins to deviate from the exponential baseline. For the subset of observations with sufficiently constrained timing properties, the characteristic frequencies of the broadband noise and low-frequency QPOs generally shift toward lower values as the inferred radius increases. This timing evolution is qualitatively consistent with the expansion of the hot inner flow.

    \item The inferred onset and pace of disk recession vary substantially among outbursts. The fiducial break scale $\dot{m}_{\rm break}$ spans a broad range, while the recession slope $\gamma$ shows no statistically significant correlation with $\dot{m}_{\rm break}$ in the present sample. This indicates that the pace of recession is not determined solely by the accretion rate scale at which the thermal luminosity deficit begins. The observed diversity likely reflects additional source- and outburst-dependent factors, such as irradiation efficiency, disk size, evaporation efficiency, magnetic configuration, and the accretion history of the outburst.
\end{enumerate}

These results support a picture in which the late decay of BHXRB outbursts is accompanied by the recession of the optically thick disk and the growth of a hot inner flow. The luminosity-deficit method provides a practical way to systematically trace this evolution when direct spectral measurements of the disk inner radius become unreliable.

\begin{acknowledgments}

This work made use of data and software provided by the High Energy Astrophysics Science Archive Research Center (HEASARC), which is a service of the Astrophysics Science Division at NASA/GSFC and the High Energy Astrophysics Division of the Smithsonian Astrophysical Observatory. The data for this analysis were obtained by the Rossi X-ray Timing Explorer (RXTE), sourced from the NASA HEASARC archive. B.Y.\ is supported by NSFC grants 12322307, 12273026, and 12361131579; supported by “the Fundamental Research Funds for the Central Universities”; Xiaomi Foundation / Xiaomi Young Talents Program; The data analysis in this paper has been done on the supercomputing system in the Supercomputing Center of Wuhan University. Z.Y. is supported by Natural Science Foundation of China (NSFC) grants 12373049. AAZ acknowledges support from the Polish National Science Center through grant 2023/48/Q/ST9/00138. A.M.\ acknowledges support from the National Science Centre, Poland (Narodowe Centrum Nauki, NCN), under grant No.\ 2018/31/G/ST9/03224.

\end{acknowledgments}

\clearpage

\restartappendixnumbering
\appendix

\section{Representative Spectral Fits and Timing Properties}
\label{sec:spec_timing_appendix}

\begin{figure*}[!htbp]
\centering
\includegraphics[width=0.8\textwidth]{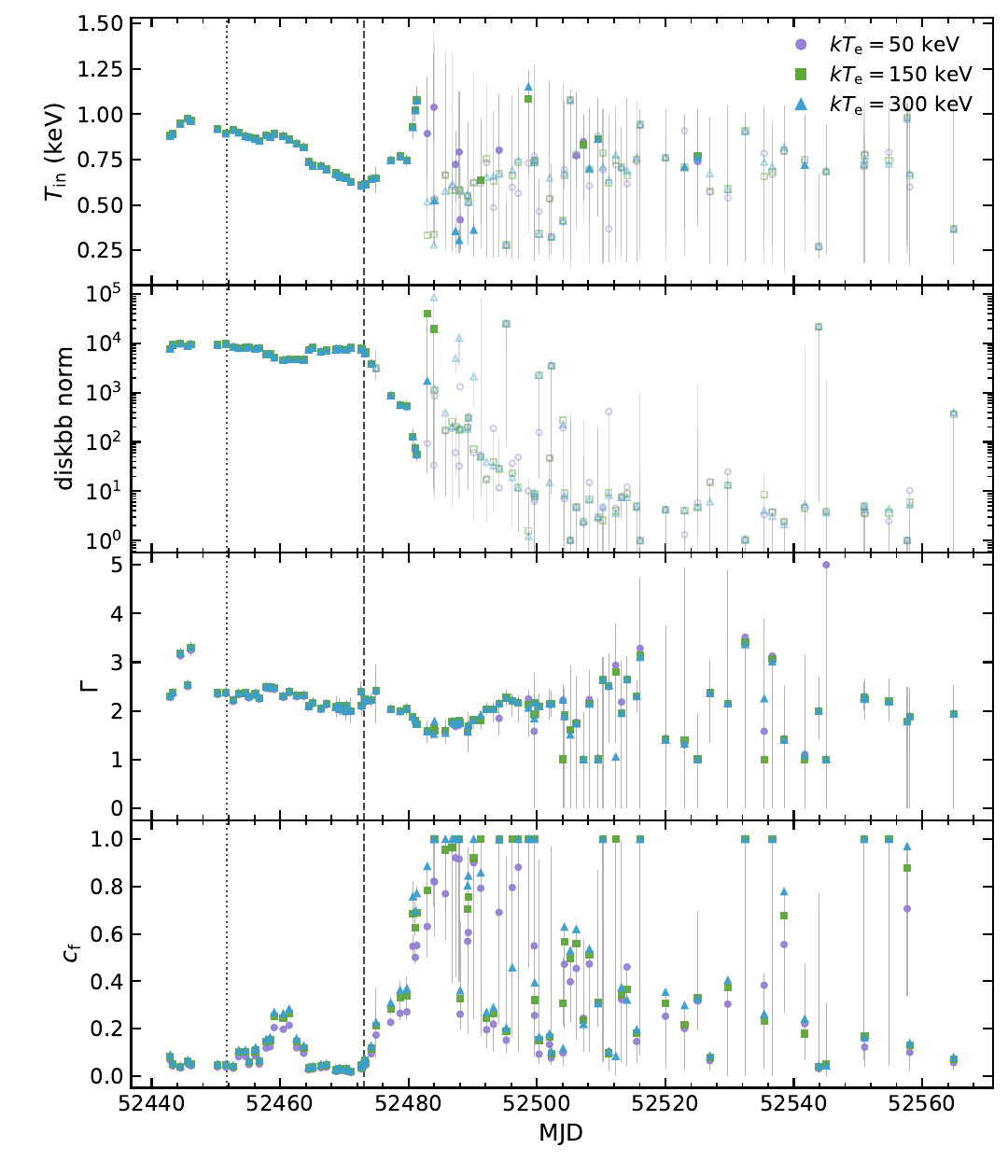}
\caption{
Key spectral parameters of 4U~1543$-$47 during its 2002 outburst. Different markers denote different fixed values of $kT_{\rm e}$. The dotted and dashed vertical lines mark the beginning and end of the exponential decay interval determined from the MCMC fitting. The onset of the inferred disk recession in our model is accompanied by an increase in the covering fraction $c_f$ of \texttt{thcomp}. In the panels showing disk parameters, hollow markers indicate epochs for which the disk parameters are weakly constrained.
}
\label{fig:4u1543_paras}
\end{figure*}

Figure~\ref{fig:4u1543_paras} shows the evolution of key spectral parameters of 4U~1543$-$47 obtained with different fixed values of $kT_{\rm e}$. This comparison shows that the fitted disk parameters, photon index, and covering fraction are not significantly affected by the assumed electron temperature in this work. The beginning and end of the fitted exponential decay interval are also marked for reference.

\begin{figure*}[!htbp]
\centering
\includegraphics[width=0.6\textwidth]{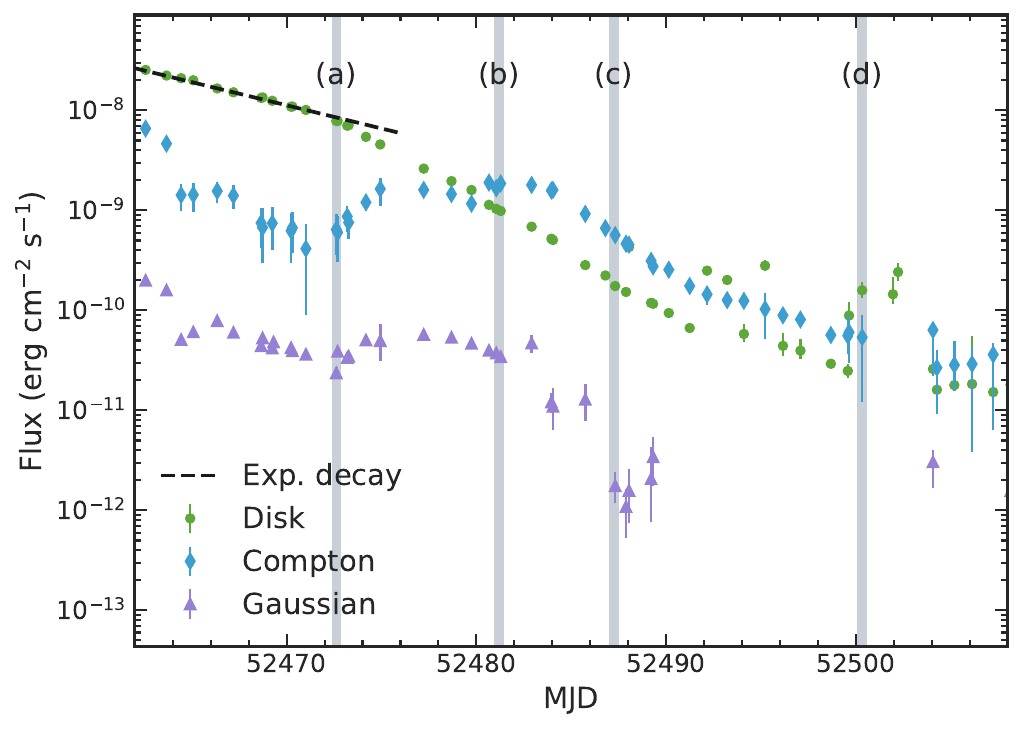}
\caption{
Model flux evolution of 4U~1543$-$47 during the soft-to-hard transition and subsequent hard-state decay. Green circles, blue diamonds, and purple triangles denote the intrinsic disk, Comptonized, and Gaussian Fe~K$\alpha$ line components, respectively. The grey shaded regions mark the four representative epochs shown in Figure~\ref{fig:spec_pds_evolution}. The dashed line shows the fitted exponential decay baseline.
}
\label{fig:flux_evolution}
\end{figure*}

\begin{figure*}[!htbp]
\centering
\includegraphics[width=1.0\textwidth]{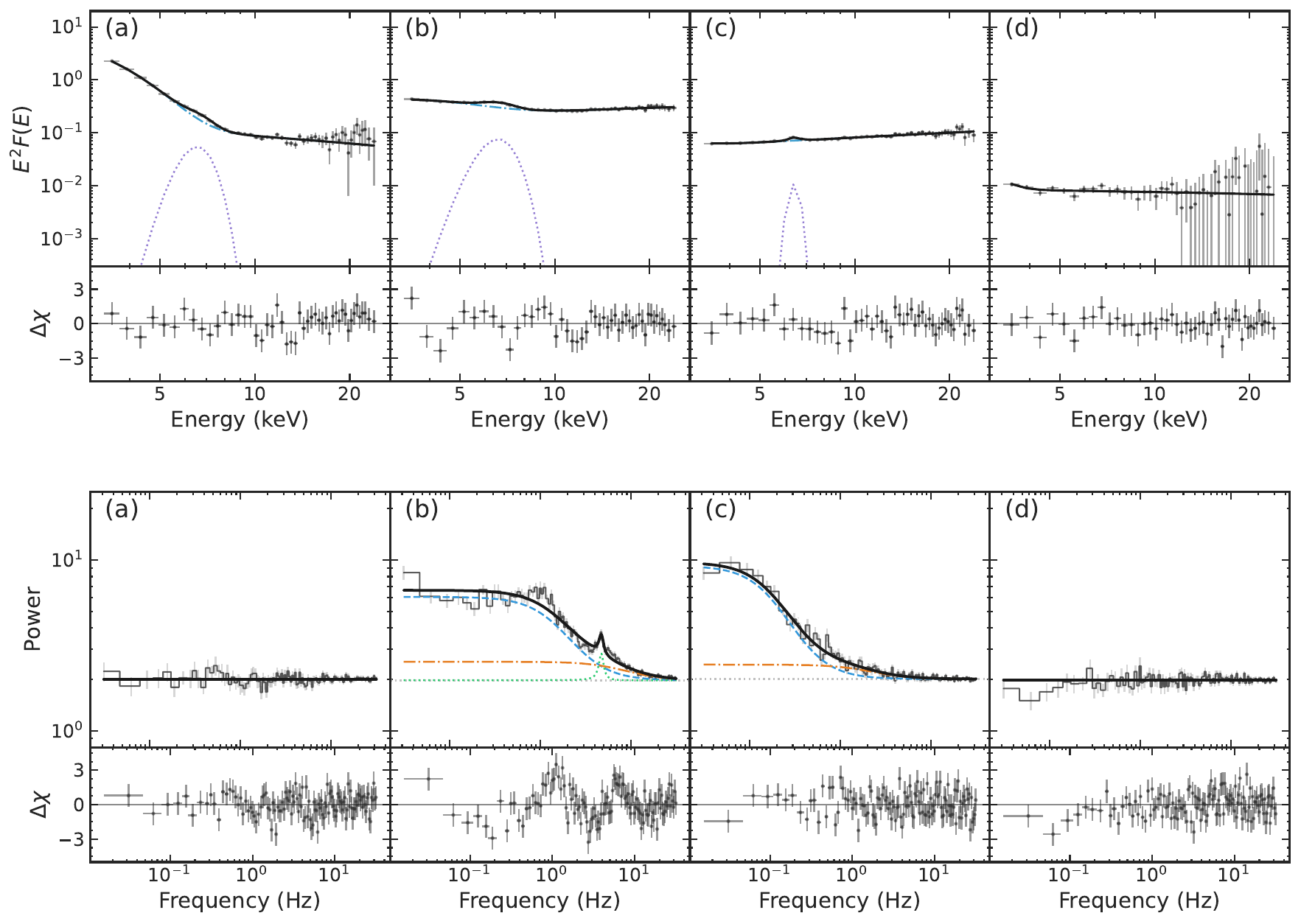}
\caption{
Energy spectra and power density spectra (PDS) for the four representative epochs marked in Figure~\ref{fig:flux_evolution}. Each column corresponds to one epoch.
\textbf{Upper panels}: Energy spectral decomposition and residuals. The \texttt{thcomp}$\otimes$\texttt{diskbb} and \texttt{gaussian} components are shown by blue dash-dotted and purple dotted lines, respectively. The total model is shown as a black solid line. The lower sub-panels show the residuals in units of $\Delta\chi=({\rm data}-{\rm model})/{\rm error}$.
\textbf{Lower panels}: The PDS are shown in the Leahy normalization for visualization. The grey dotted lines indicate the fitted Poisson noise level around 2. The low- and high-frequency band-limited noise components are denoted by blue dashed and orange dash-dotted lines, respectively. In panel (b), the green dotted line marks a low-frequency QPO component. The black solid lines represent the total Lorentzian model. The lower sub-panels show the residuals in units of $\Delta\chi$.
}
\label{fig:spec_pds_evolution}
\end{figure*}

Figure~\ref{fig:flux_evolution} shows the model flux evolution of 4U~1543$-$47 during the soft-to-hard transition and subsequent hard-state decay. This source is used as a representative example because its disk luminosity shows a clear departure from the initial decay trend, accompanied by an enhancement of the Comptonized component. The four shaded intervals in Figure~\ref{fig:flux_evolution} mark the epochs for which the corresponding spectral decompositions and power density spectra (PDS) are shown in Figure~\ref{fig:spec_pds_evolution}. These examples illustrate the transition from a disk-dominated spectrum with weak intrinsic variability to a Compton-dominated phase with enhanced broadband noise and low-frequency QPO-like features.

The representative PDS in Figure~\ref{fig:spec_pds_evolution} are shown in the Leahy normalization for visualization. In this normalization, the nominal Poisson noise level is 2 \citep{1983ApJ...266..160L}, providing a convenient reference for identifying intrinsic source variability. The Lorentzian curves overlaid in this figure were refitted to the Leahy-normalized PDS for display purposes only. The quantitative timing parameters reported in this work were instead obtained from the fractional-rms-normalized PDS after Poisson-noise subtraction, as described in Section~\ref{sec:data}.

\section{MCMC Fits for the Exponential Baseline}
\label{sec:fit_exp_baseline}

To characterize the soft state viscous decay of the disk emission, we fitted the disk luminosity with an exponential function over a statistically determined temporal interval. The boundaries of this interval were treated as free parameters rather than fixed manually.

Within a continuous interval $[t_{\rm start}, t_{\rm end}]$, the expected disk luminosity is modeled as
\begin{equation}
L_{\rm d}^{\rm exp}(t)
=
A \exp\left[-\frac{t-t_{\rm start}}{\tau}\right],
\label{eq:exp_decay_appendix}
\end{equation}
where $A$ is the luminosity normalization at $t_{\rm start}$, $\tau$ is the e-folding decay timescale, and $t_{\rm start}$ and $t_{\rm end}$ denote the start and end times of the exponential baseline, respectively.

To allow for moderate intrinsic fluctuations around the smooth decay trend, we added a fractional intrinsic scatter term, $\sigma_{\rm int}=0.05$, in quadrature with the observational uncertainty. The total uncertainty for the $i$-th observation is therefore
\begin{equation}
\sigma_{i,\rm tot}^{2}
=
\sigma_{i,\rm obs}^{2}
+
\left(\sigma_{\rm int} m_i\right)^{2},
\label{eq:exp_total_uncertainty}
\end{equation}
where $m_i=L_{\rm d}^{\rm exp}(t_i)$ is the model prediction. When asymmetric observational uncertainties are available, the uncertainty on the appropriate side of the model residual was used.

We adopted a piecewise Gaussian likelihood. Observations within $[t_{\rm start},t_{\rm end}]$ directly constrain the exponential baseline, whereas observations outside this interval are retained with reduced statistical weight:
\begin{equation}
\mathcal{L}_{i} =
\begin{cases}
\mathcal{N}\left(y_i; m_i, \sigma_{i,\rm tot}^{2}\right), & t_i \in [t_{\rm start},t_{\rm end}], \\
\mathcal{N}\left(y_i; m_i, f_{\rm out}^{2}\sigma_{i,\rm tot}^{2}\right), & t_i \notin [t_{\rm start},t_{\rm end}],
\end{cases}
\label{eq:exp_piecewise_likelihood}
\end{equation}
where $y_i$ is the observed disk luminosity and $f_{\rm out}=5$. The down-weighted points outside the fitted interval help preserve the global decay trend while allowing the exponential baseline to be determined primarily by the smooth soft state segment. The fitted parameter set is $\boldsymbol{\theta}_{\rm exp} = \{{A,\tau,t_{\rm start},t_{\rm end}}\}$.

The posterior distribution was sampled with \texttt{emcee} \citep{2013PASP..125..306F}. We used 32 walkers and 5000 steps per walker, discarding the first 1000 steps as burn-in. The reported values correspond to the posterior medians, and the uncertainties are given by the 16th and 84th percentiles. The resulting parameters are listed in Table~\ref{tab:fit_results}.

\begin{deluxetable}{lcccc}
\tabletypesize{\footnotesize}
\tablewidth{0pt}
\tablecaption{Best-fit parameters of the soft-state exponential decay baseline\label{tab:fit_results}}
\tablehead{
    \colhead{Source (Outburst Year)} & \colhead{$A$} & \colhead{$\tau$} & \colhead{$t_{\rm start}$} & \colhead{$t_{\rm end}$} \\
    \colhead{} & \colhead{($10^{37}\,\mathrm{erg\,s^{-1}}$)} & \colhead{(days)} & \colhead{(MJD)} & \colhead{(MJD)}
}
\startdata
4U 1543$-$47 (2002) & $59.86^{+7.37}_{-6.57}$ & $9.4 \pm 0.2$ & $52451.7^{+1.1}_{-1.2}$ & $52473.1^{+0.7}_{-1.3}$ \\
GRO J1655$-$40 (1996) & $6.20^{+0.37}_{-0.39}$ & $89.5^{+6.5}_{-7.3}$ & $50590.8^{+5.5}_{-6.6}$ & $50672.7^{+21.2}_{-20.2}$ \\
GRO J1655$-$40 (2005) & $3.22^{+0.18}_{-0.13}$ & $18.5 \pm 0.5$ & $53599.3^{+0.8}_{-0.9}$ & $53628.0^{+0.8}_{-1.1}$ \\
GX 339$-$4 (2002) & $5.90^{+0.43}_{-0.62}$ & $23.3^{+1.0}_{-1.5}$ & $52662.9^{+3.0}_{-1.6}$ & $52693.2^{+12.3}_{-1.8}$ \\
GX 339$-$4 (2004) & $3.77^{+0.13}_{-0.14}$ & $72.0^{+10.1}_{-8.6}$ & $53410.8^{+3.5}_{-3.2}$ & $53461.4^{+4.3}_{-5.1}$ \\
GX 339$-$4 (2006) & $14.78^{+0.43}_{-0.45}$ & $30.7^{+0.3}_{-0.4}$ & $54153.1^{+1.1}_{-0.9}$ & $54230.9^{+0.7}_{-3.0}$ \\
GX 339$-$4 (2010) & $2.37^{+0.13}_{-0.15}$ & $33.7^{+2.4}_{-2.1}$ & $55562.5^{+2.4}_{-1.0}$ & $55593.4^{+1.0}_{-2.1}$ \\
H1743$-$322 (2002) & $10.53^{+0.63}_{-0.68}$ & $17.7^{+0.7}_{-0.6}$ & $52910.7^{+1.3}_{-1.0}$ & $52935.3^{+0.6}_{-1.4}$ \\
H1743$-$322 (2004) & $9.24^{+2.18}_{-2.12}$ & $13.5 \pm 0.4$ & $53265.5^{+3.6}_{-2.9}$ & $53294.7^{+1.5}_{-3.3}$ \\
H1743$-$322 (2005) & $9.13^{+0.59}_{-0.97}$ & $17.4^{+1.4}_{-1.3}$ & $53602.1^{+1.7}_{-1.0}$ & $53620.9^{+4.5}_{-4.9}$ \\
H1743$-$322 (2008) & $4.60^{+0.26}_{-0.32}$ & $11.7^{+0.8}_{-0.7}$ & $54483.8^{+0.7}_{-0.8}$ & $54497.4^{+0.9}_{-1.1}$ \\
H1743$-$322 (2009) & $10.51^{+1.53}_{-0.79}$ & $18.2^{+1.1}_{-1.0}$ & $54987.9^{+2.1}_{-2.6}$ & $55011.5^{+3.4}_{-4.4}$ \\
H1743$-$322 (2010) & $7.20^{+0.49}_{-0.35}$ & $19.1 \pm 0.7$ & $55431.6^{+1.1}_{-0.9}$ & $55455.4^{+0.9}_{-0.7}$ \\
H1743$-$322 (2011) & $6.91^{+0.47}_{-0.44}$ & $12.7 \pm 1.3$ & $55676.0^{+1.7}_{-0.6}$ & $55686.0 \pm 1.2$ \\
MAXI J1543$-$564 (2011)\tablenotemark{a} & $2.73 \pm 0.05$ & $55.3^{+1.5}_{-1.4}$ & $55731.8^{+1.0}_{-0.5}$ & $55824.5^{+6.3}_{-5.8}$ \\
MAXI J1659$-$152 (2010) & $4.07^{+0.15}_{-0.33}$ & $18.6^{+1.1}_{-1.0}$ & $55484.7^{+1.3}_{-0.3}$ & $55500.7 \pm 0.4$ \\
SLX 1746$-$331 (2003) & $2.23^{+0.13}_{-0.18}$ & $67.1^{+8.0}_{-6.3}$ & $52788.1^{+3.4}_{-2.5}$ & $52828.2^{+2.4}_{-3.7}$ \\
SLX 1746$-$331 (2007)\tablenotemark{a} & $3.63^{+0.20}_{-0.28}$ & $9.1 \pm 0.4$ & $54384.5^{+0.9}_{-0.5}$ & $54397.4^{+1.8}_{-1.7}$ \\
Swift J1539.2$-$6227 (2008) & $0.58^{+0.03}_{-0.04}$ & $25.9^{+1.1}_{-0.9}$ & $54899.4^{+1.4}_{-1.1}$ & $54936.0^{+0.5}_{-1.7}$ \\
Swift J1842.5$-$1124 (2008)\tablenotemark{a} & $2.13^{+0.07}_{-0.08}$ & $29.9 \pm 0.6$ & $54734.3^{+1.4}_{-0.7}$ & $54824.5^{+17.4}_{-17.3}$ \\
XTE J1550$-$564 (1998) & $3.85^{+0.42}_{-0.45}$ & $10.3^{+0.3}_{-0.2}$ & $51266.9^{+1.5}_{-1.4}$ & $51304.2 \pm 0.7$ \\
XTE J1550$-$564 (2000) & $5.84^{+0.45}_{-0.77}$ & $14.7^{+1.7}_{-1.1}$ & $51661.4^{+2.0}_{-0.9}$ & $51672.6^{+0.4}_{-0.9}$ \\
XTE J1650$-$500 (2001) & $1.32^{+0.11}_{-0.13}$ & $26.8^{+1.1}_{-1.0}$ & $52194.5^{+3.1}_{-2.6}$ & $52230.8^{+0.8}_{-1.1}$ \\
XTE J1652$-$453 (2009)\tablenotemark{b} & $9.62^{+0.28}_{-0.80}$ & $22.0 \pm 0.3$ & $55028.7^{+2.1}_{-0.6}$ & $55096.7^{+2.3}_{-2.8}$ \\
XTE J1720$-$318 (2003) & $3.44^{+0.15}_{-0.22}$ & $22.0^{+1.2}_{-0.9}$ & $52691.3^{+1.7}_{-1.5}$ & $52729.6^{+3.7}_{-4.2}$ \\
XTE J1748$-$288 (1998)\tablenotemark{a} & $15.27^{+0.83}_{-1.24}$ & $19.1^{+1.1}_{-1.0}$ & $50977.8^{+2.6}_{-1.3}$ & $51008.5^{+2.6}_{-4.5}$ \\
XTE J1752$-$223 (2009) & $7.71^{+0.14}_{-0.13}$ & $38.9^{+1.0}_{-1.3}$ & $55218.6^{+0.4}_{-0.3}$ & $55272.2^{+5.3}_{-2.0}$ \\
XTE J1817$-$330 (2006) & $1.82 \pm 0.08$ & $24.1^{+0.7}_{-0.6}$ & $53842.1^{+1.3}_{-1.0}$ & $53889.0^{+1.9}_{-4.9}$ \\
XTE J1818$-$245 (2005)\tablenotemark{a} & $0.96 \pm 0.05$ & $20.9 \pm 0.4$ & $53630.1^{+1.1}_{-1.2}$ & $53709.3^{+7.3}_{-6.9}$ \\
XTE J1859$+$226 (1999) & $9.79^{+1.33}_{-1.18}$ & $26.7^{+1.0}_{-1.2}$ & $51534.0^{+3.1}_{-4.2}$ & $51604.9^{+6.1}_{-10.1}$ \\
\enddata
\tablenotetext{a}{Excluded from the 24-outburst recession sample owing to insufficient post-$t_{\rm end}$ coverage with a constrained disk luminosity deficit (fewer than two measurements with $R_{\rm tr}/R_{\rm ISCO} > 1.01$).}
\tablenotetext{b}{Excluded from the 24-outburst recession sample because all post-$t_{\rm end}$ disk luminosities lie at or above the extrapolated exponential baseline, yielding only lower limits ($R_{\rm tr} = R_{\rm ISCO}$).}
\tablecomments{Reported values and uncertainties correspond to the posterior medians and 16th--84th percentiles from the MCMC fits (Appendix~\ref{sec:fit_exp_baseline}). Where the upper and lower uncertainties are equal after rounding, they are written in $\pm$ form.}
\end{deluxetable}

\onecolumngrid
\section{Individual Light Curves and Supplementary Diagnostics}
\label{sec:individual}

\begin{figure*}[tbp]
    \centering
    \includegraphics[width=\textwidth]{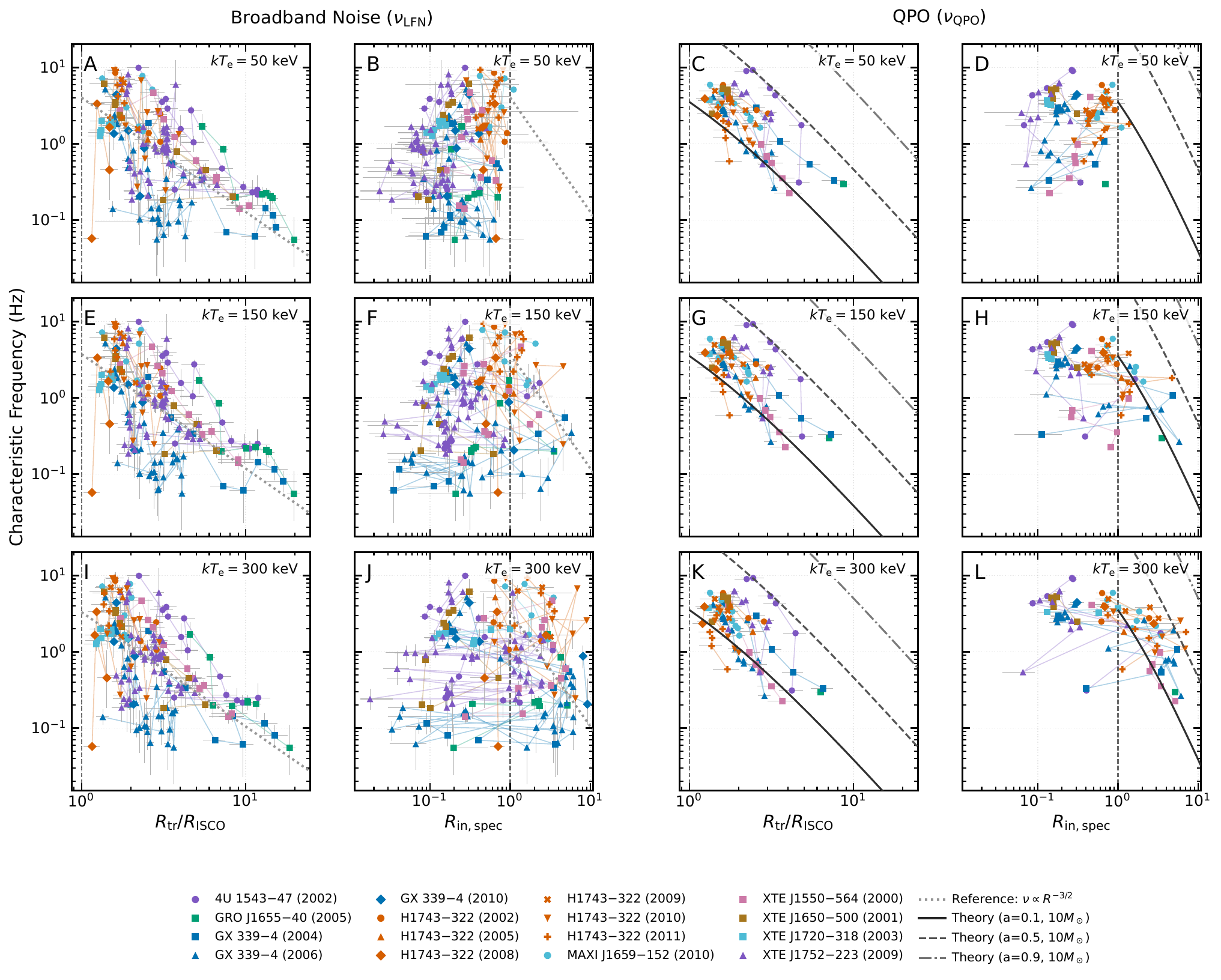}
    \caption{
    Comparison of characteristic timing frequencies against the luminosity-derived truncation radius ratio $R_{\rm tr}/R_{\rm ISCO}$ and the relative spectral inner radius $R_{\rm in,spec} \equiv \sqrt{N_{\rm diskbb}/N_{\rm diskbb,soft}}$ for three fixed coronal electron temperatures ($kT_{\rm e} = 50$, $150$, and $300\,\mathrm{keV}$, from top to bottom). The left two columns show the broadband-noise frequency ($\nu_{\rm LFN}$) with a $\nu \propto R^{-3/2}$ visual reference line (dotted), and the right two columns show the QPO frequency ($\nu_{\rm QPO}$) with rigid Lense--Thirring precession model curves for $M_{\rm BH} = 10\,M_\odot$ and representative spins $a_* = 0.1, 0.5, 0.9$ \citep{2009MNRAS.397L.101I}.
    }
    \label{fig:cosistency_check}
\end{figure*}

\begin{figure*}[tbp]
    \centering
    \includegraphics[height=0.84\textheight,keepaspectratio]{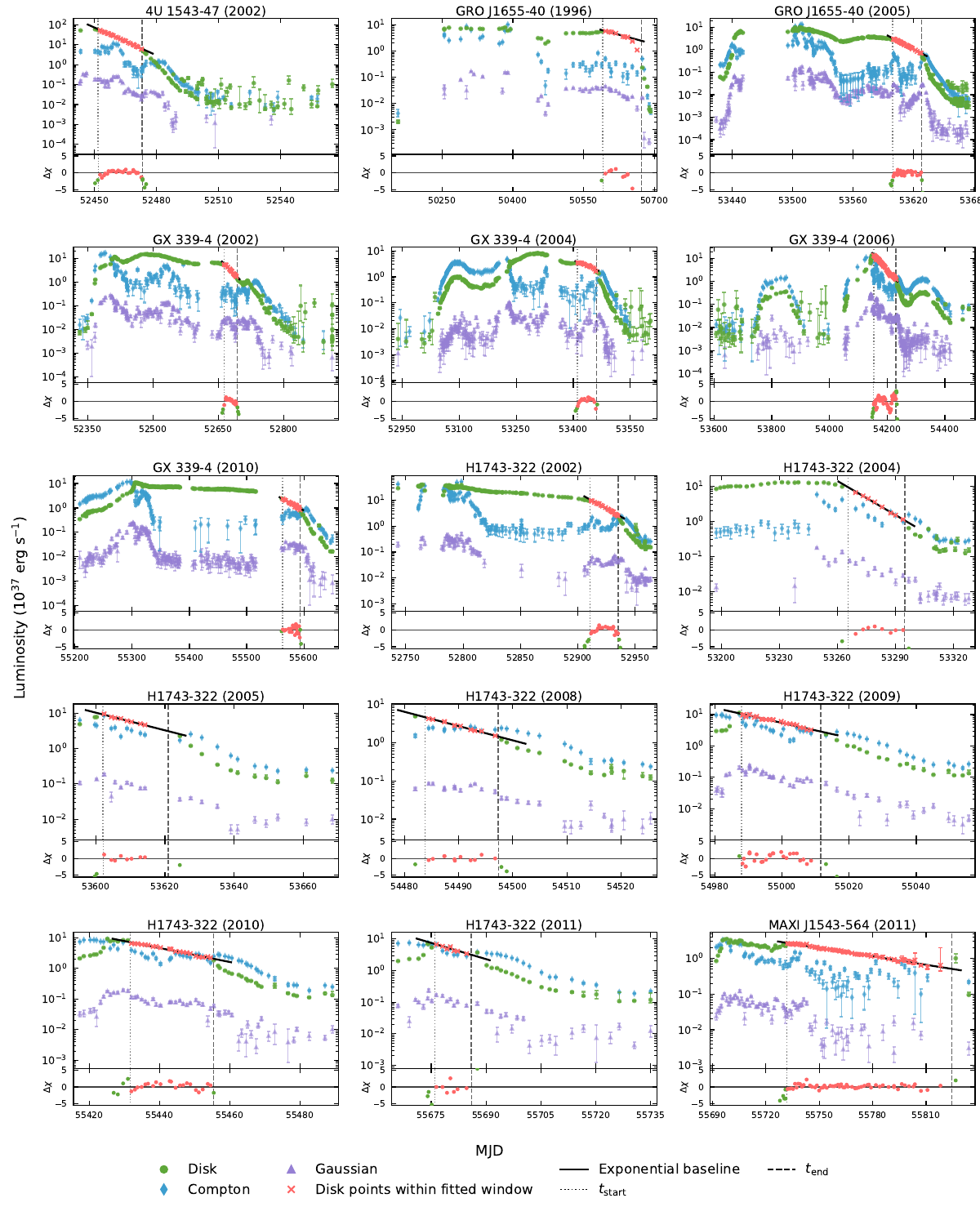}
    \caption{
    Individual exponential decay fits for the selected outbursts. Red points indicate the fitted disk-luminosity data, black curves show the best-fitting exponential baselines, and vertical dotted/dashed lines mark $t_{\rm start}$/$t_{\rm end}$.
    }
    \label{fig:decay_fits_page1}
\end{figure*}

\begin{figure*}[tbp]
    \centering
    \includegraphics[height=0.84\textheight,keepaspectratio]{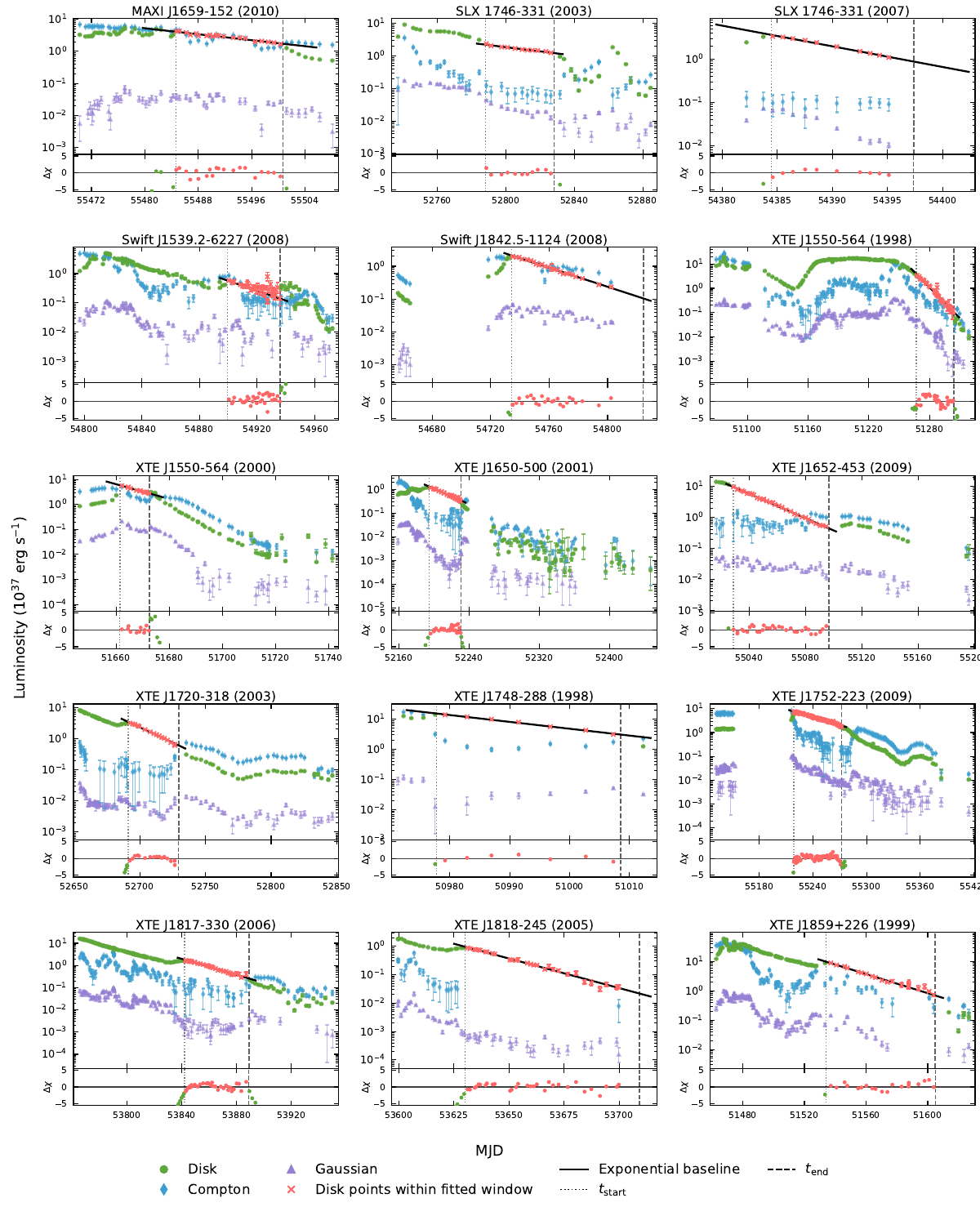}
    \caption{
    Same as Figure~\ref{fig:decay_fits_page1}, for the remaining selected outbursts.
    }
    \label{fig:decay_fits_page2}
\end{figure*}

Figure~\ref{fig:cosistency_check} presents the sample-wide comparison of characteristic timing frequencies against the luminosity-derived $R_{\rm tr}/R_{\rm ISCO}$ and the relative spectral inner radius $R_{\rm in,spec}$ discussed in Section~\ref{subsec:spectral-radius}. Figures~\ref{fig:decay_fits_page1} and \ref{fig:decay_fits_page2} show the individual disk luminosity decay fits for the outbursts included in the exponential baseline analysis. In each panel, the full luminosity evolution is shown for the disk, Comptonized, and Gaussian components. The disk luminosity points used to determine the exponential baseline are highlighted in red, while the black curve shows the best-fitting exponential model over the fitted decay interval. The vertical dotted and dashed lines mark the fitted start and end times of the exponential window, $t_{\rm start}$ and $t_{\rm end}$, respectively. The lower sub-panel gives the residuals in the local fitting region. These figures provide a source-by-source view of the baseline selection and the subsequent disk luminosity deficit used to construct the recession tracer.

\begin{figure*}[tbp]
    \centering
    \includegraphics[width=\textwidth]{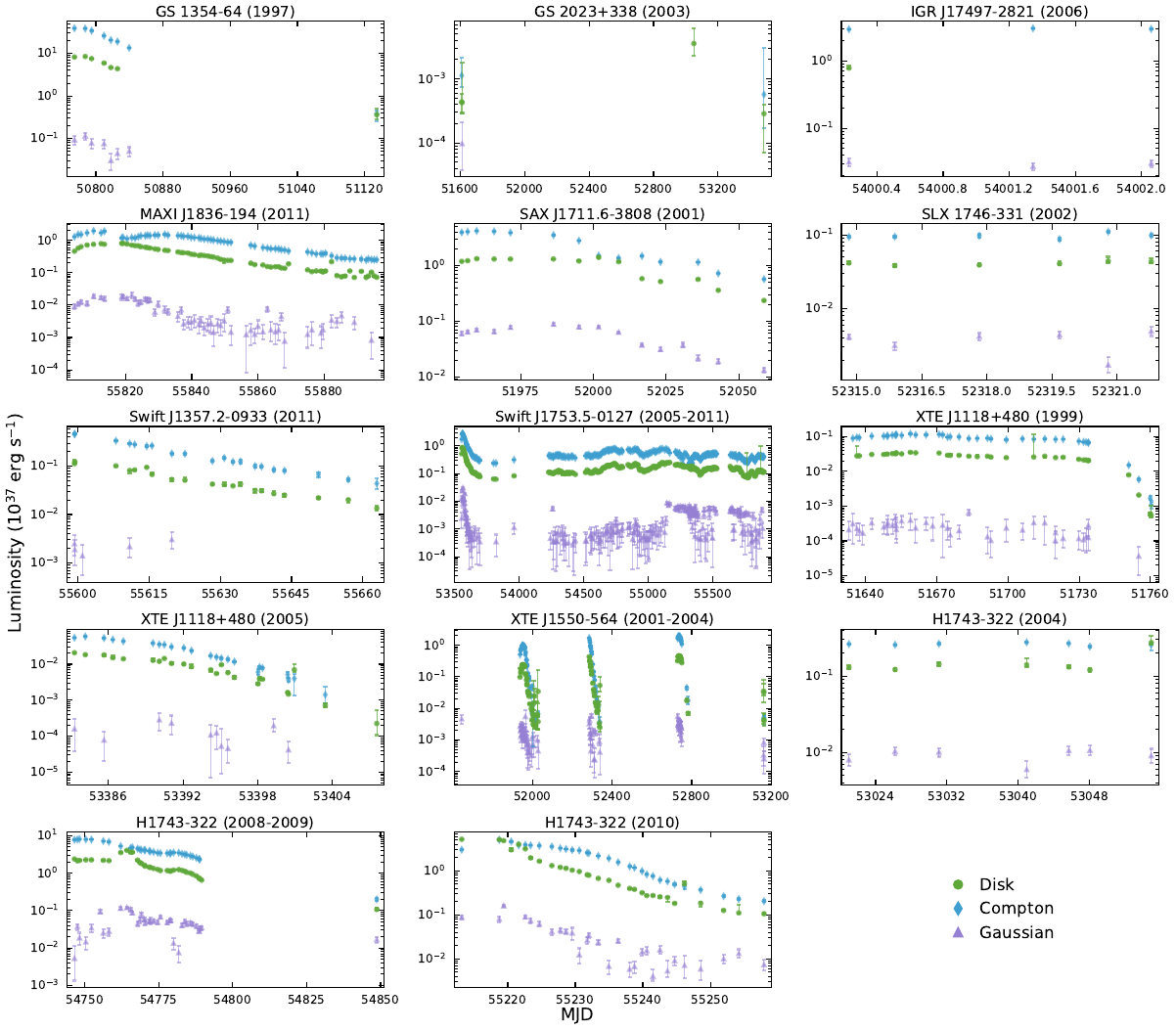}
    \caption{
    Luminosity evolution of outbursts or decay episodes not included in the exponential-decay fitting sample.
    }
    \label{fig:unfitted_outbursts}
\end{figure*}

For comparison, Figure~\ref{fig:unfitted_outbursts} shows the outbursts or decay episodes that were inspected but not included in the exponential decay fitting sample. These cases were excluded because they do not provide a robust disk-dominated soft state exponential baseline, either due to insufficient soft state coverage, hard-state-only observations, failed-outburst behavior, or weak and irregular disk emission.

\begin{deluxetable}{llcccccc}
\tablewidth{0pt}
\tablecaption{Constrained disk truncation radius ratios ($R_{\rm tr}/R_{\rm ISCO}$) during post-soft-state decays\label{tab:rtr_values}}
\tablehead{
    \colhead{Source} & \colhead{Year} & \colhead{ObsID} & \colhead{MJD} & \colhead{$t - t_{\rm end}$} & \colhead{$R_{\rm tr}/R_{\rm ISCO}$} & \colhead{$\dot{m}_{\rm d}^{\rm exp}$} & \colhead{Flag} \\
    \colhead{} & \colhead{} & \colhead{} & \colhead{} & \colhead{(days)} & \colhead{} & \colhead{} & \colhead{}
}
\startdata
4U 1543$-$47 & 2002 & 70133-01-29-00 & 52474.18 &  1.08 & $1.32 \pm 0.24$  & $0.0460 \pm 0.0082$ & 1 \\
4U 1543$-$47 & 2002 & 70133-01-30-00 & 52474.93 &  1.83 & $1.45 \pm 0.30$  & $0.0424 \pm 0.0076$ & 0 \\
4U 1543$-$47 & 2002 & 70133-01-31-01 & 52477.23 &  4.12 & $1.99 \pm 0.37$  & $0.0332 \pm 0.0060$ & 1 \\
4U 1543$-$47 & 2002 & 70124-02-01-00 & 52478.69 &  5.58 & $2.27 \pm 0.42$  & $0.0284 \pm 0.0052$ & 1 \\
4U 1543$-$47 & 2002 & 70124-02-02-00 & 52479.74 &  6.64 & $2.49 \pm 0.47$  & $0.0254 \pm 0.0047$ & 1 \\
4U 1543$-$47 & 2002 & 70124-02-03-00 & 52480.67 &  7.56 & $3.17 \pm 0.59$  & $0.0230 \pm 0.0043$ & 1 \\
4U 1543$-$47 & 2002 & 70128-01-01-01 & 52481.05 &  7.95 & $3.34 \pm 0.62$  & $0.0221 \pm 0.0041$ & 1 \\
4U 1543$-$47 & 2002 & 70128-01-01-00 & 52481.28 &  8.18 & $3.42 \pm 0.63$  & $0.0215 \pm 0.0040$ & 1 \\
4U 1543$-$47 & 2002 & 70124-02-05-00 & 52482.91 &  9.81 & $4.12 \pm 0.79$  & $0.0181 \pm 0.0034$ & 0 \\
4U 1543$-$47 & 2002 & 70124-02-06-00 & 52483.95 & 10.85 & $4.87 \pm 0.92$  & $0.0162 \pm 0.0031$ & 0 \\
4U 1543$-$47 & 2002 & 70124-02-07-00 & 52484.04 & 10.93 & $4.97 \pm 0.94$  & $0.0161 \pm 0.0030$ & 0 \\
4U 1543$-$47 & 2002 & 70124-02-08-00 & 52485.74 & 12.63 & $7.39 \pm 1.45$  & $0.0134 \pm 0.0026$ & 0 \\
4U 1543$-$47 & 2002 & 70124-02-09-00 & 52486.80 & 13.70 & $8.38 \pm 1.62$  & $0.0119 \pm 0.0023$ & 0 \\
4U 1543$-$47 & 2002 & 70128-01-02-00 & 52487.31 & 14.21 & $10.11 \pm 1.95$ & $0.0113 \pm 0.0022$ & 0 \\
4U 1543$-$47 & 2002 & 70128-01-02-01 & 52487.89 & 14.78 & $10.90 \pm 2.11$ & $0.0106 \pm 0.0021$ & 0 \\
\enddata
\tablecomments{Table~\ref{tab:rtr_values} is published in its entirety in the machine-readable format (for all 24 outbursts). A portion is shown here for guidance regarding its form and content. Flag $= 1$ indicates well-constrained \texttt{diskbb} spectral parameters, while Flag $= 0$ indicates weakly constrained disk measurements.}
\end{deluxetable}

\clearpage

\bibliography{main}{}
\bibliographystyle{aasjournalv7}

\end{document}